\documentclass[apj,twocolumn]{openjournal}
\usepackage{graphicx}
\usepackage{xcolor}
\usepackage{amsmath}
\usepackage{amsfonts}
\usepackage{amssymb}
\usepackage{upgreek}
\usepackage[pdfborder={0 0 0}]{hyperref}
\usepackage{txfonts}
\usepackage{lipsum} 
\newcommand{\orcidauthor}[3]{\author{\href{http://orcid.org/#1}{#2 \openin1 Orcid-ID.png \ifeof1 \else \hskip2pt\includegraphics[width=9pt]{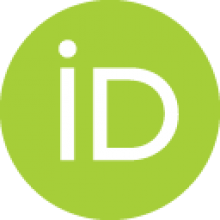}\fi}$^{#3}$}}

\graphicspath{ {./figs/} }

\begin{document}
\title{Understanding the Neutron Star Population with the SKAO telescopes}

\orcidauthor{0000-0002-2034-2986}{L. Levin}{1}
\orcidauthor{0000-0001-8640-8186}{M. Bagchi}{2,3}
\orcidauthor{0000-0002-8265-4344}{M. Burgay}{4}
\orcidauthor{0000-0001-9434-3837}{A. T. Deller}{5}
\orcidauthor{0000-0002-6558-1681}{V. Graber}{6}
\orcidauthor{0000-0003-2145-1022}{A. Igoshev}{7}
\orcidauthor{0000-0002-4175-2271}{M. Kramer}{8,1}
\orcidauthor{0000-0003-1301-966X}{D. Lorimer}{9}
\orcidauthor{0000-0003-2317-9747}{B. Posselt}{10}
\orcidauthor{0000-0003-4038-8065}{T. Prabu}{11}
\orcidauthor{0000-0002-8043-6909}{K. Rajwade}{10}
\orcidauthor{0000-0003-2177-6388}{N. Rea}{12,13}
\orcidauthor{0000-0001-9242-7041}{B. Stappers}{1}
\orcidauthor{0000-0002-3865-7265}{T. M. Tauris}{14,8}
\orcidauthor{0000-0003-2122-4540}{P. Weltevrede}{1}
\author{The SKAO Pulsar Science Working Group}

\affiliation{$^1$Jodrell Bank Centre for Astrophysics, Department of Physics and Astronomy, The University of Manchester, Manchester M13 9PL, UK}
\affiliation{$^2$The Institute of Mathematical Sciences, Taramani, Chennai 600113, India}
\affiliation{$^3$Homi Bhabha National Institute, Training School Complex, Anushakti Nagar, Mumbai 400094, India}
\affiliation{$^4$INAF – Osservatorio Astronomico di Cagliari, Via della Scienza 5, I-09047 Selargius, (CA), Italy}
\affiliation{$^5$Centre for Astrophysics and Supercomputing, Swinburne University of Technology, Hawthorn, VIC 3122, Australia}
\affiliation{$^6$Department of Physics, Royal Holloway, University of London, Egham, TW20 0EX, UK}
\affiliation{$^7$School of Mathematics, Statistics and Physics, Newcastle University, Newcastle upon Tyne,  NE1 7RU,  UK}
\affiliation{$^{8}$Max-Planck-Institut f\"{u}r Radioastronomie, Auf dem H\"{u}gel 69, D-53121 Bonn, Germany}
\affiliation{$^9$Department of Physics and Astronomy, West Virginia University, Morgantown, WV 26506-6315, USA}
\affiliation{$^{10}$Department of Astrophysics, University of Oxford, Denys Wilkinson Building, Keble Road, Oxford OX1 3RH, UK}
\affiliation{$^{11}$Raman Research Institute, Bangalore, India}
\affiliation{$^{12}$Institute of Space Sciences (ICE-CSIC), Campus UAB, C/ de Can Magrans s/n, Cerdanyola del Vallès (Barcelona) 08193, Spain}
\affiliation{$^{13}$Institut d'Estudis Espacials de Catalunya (IEEC), Castelldefels, Spain}
\affiliation{$^{14}$Department of Materials and Production, Aalborg University, Fibigerstr{\ae}de 16, 9220 Aalborg, Denmark}

\begin{abstract}
The known population of non-accreting neutron stars is ever growing and currently consists of more than 3500 sources. 
Pulsar surveys with the SKAO telescopes will greatly increase the known population, adding radio pulsars to every subgroup in the radio-loud neutron star family. These discoveries will not only add to the current understanding of neutron star physics by increasing the sample of sources that can be studied, but will undoubtedly also uncover previously unknown types of sources that will challenge our theories of a wide range of physical phenomena. 
A broad variety of scientific studies will be made possible by a significantly increased known population of neutron stars, unravelling questions such as: How do isolated pulsars evolve with time; What is the connection between magnetars, high B-field pulsars, and the newly discovered long-period pulsars; How is a pulsar's spin-down related to its radio emission; What is the nuclear equation of state? Increasing the known numbers of pulsars in binary or triple systems may enable both larger numbers and higher precision tests of gravitational theories and general relativity, as well as probing the neutron star mass distribution.
The excellent sensitivity of the SKAO telescopes combined with the wide field of view, large numbers of simultaneous tied-array beams that will be searched in real time, wide range of observing frequencies, and the ability to form multiple sub-arrays will make the SKAO an excellent facility to undertake a wide range of neutron star research. 
In this paper, we give an overview of different types of neutron stars and discuss how the SKAO telescopes will aid in our understanding of the neutron star population.
\end{abstract}

\maketitle

\section{Introduction}

\begin{figure}[ht]
    \centering
    \includegraphics[width=1.0\linewidth]{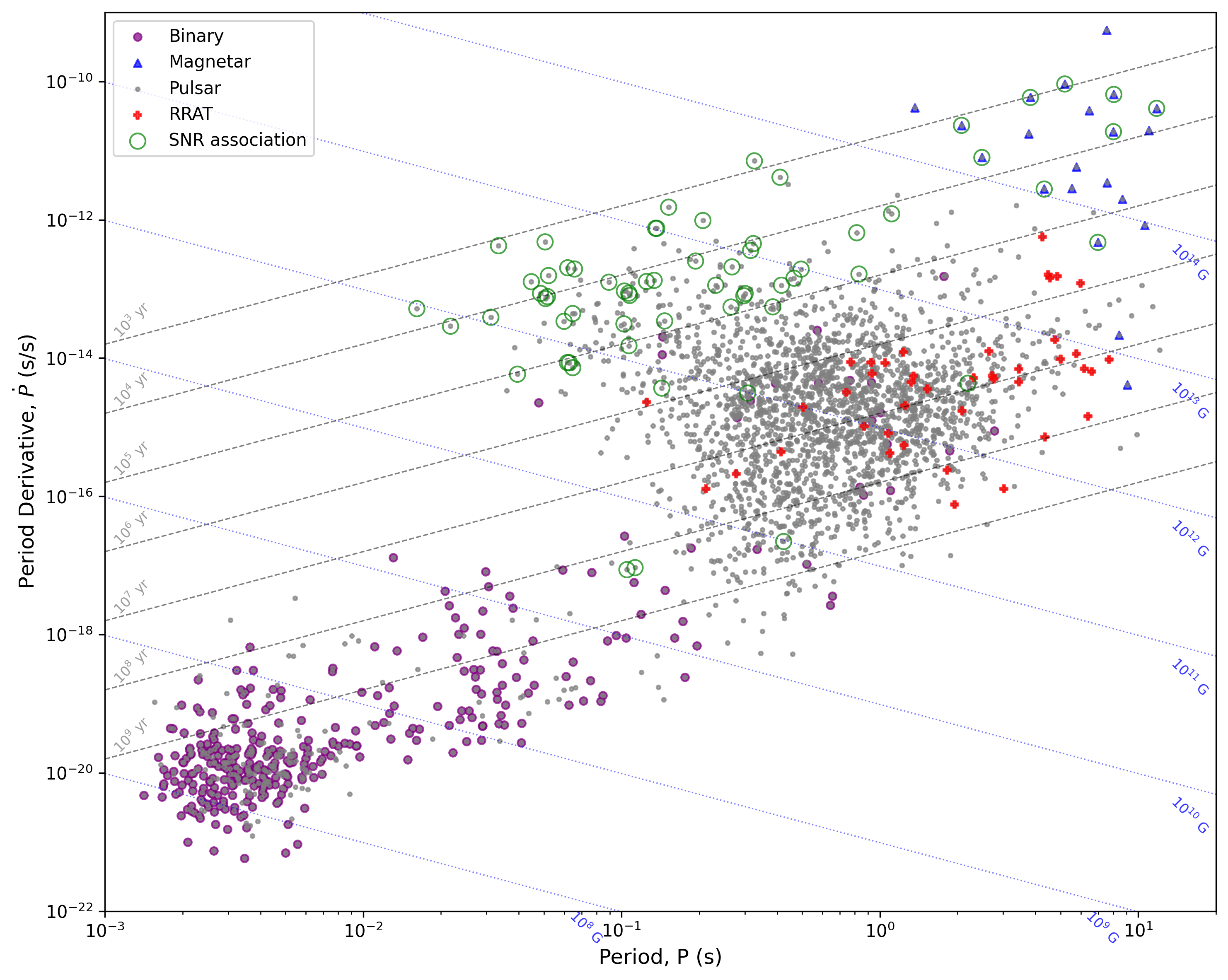}
   \caption{Distribution of 2752 pulsars in the $P-\dot{P}$ diagram. These are all the currently known pulsars that have both period and period derivative values listed. Lines of constant characteristic age and constant surface magnetic field strength are shown. 
   Data taken from the {\em ATNF Pulsar Catalogue} version 2.6.1 in June~2025 \citep[][\url{https://www.atnf.csiro.au/research/pulsar/psrcat}]{Manchester2005}.}
    \label{fig:ppdot_intro}
\end{figure}

The known population of neutron stars (NSs) has increased quickly over the last decade with the completion of new, more sensitive telescopes, such as the FAST and the MeerKAT radio telescopes, and is currently standing at over 3500 sources. The population is surprisingly diverse, with different subgroups exhibiting a variety of characteristics. 
The sources span many orders of magnitude in spin period ($P$) and period derivative ($\dot{P}$), and hence also in inferred magnetic field strength and characteristic age, as shown in the $P-\dot{P}$ diagram in Figure \ref{fig:ppdot_intro}. 
The different NS subgroups, to some extent, show different radio emission properties in terms of e.g. spectra, emission modes, variability, and intermittency, indicating different magnetospheric conditions~\citep{Oswald2025_MAG}. 
One of the important outstanding questions in the field of NS research is, if, and how all the varied subgroups fit together into one unified NS family. 

In this paper, we will describe the different subgroups of NSs, give an overview of the current state of the field, as well as discuss how observations with the SKAO telescopes will enable significant advances both in our understanding of each of the subgroups of NSs as well as of the population as a whole. 
We present these groups separately for simplicity; however, pulsars whose radio emission properties bridge the sub-populations or in other ways stand out from the rest could be the key to understanding possible evolutionary paths within the NS population. Finding new members of the existing groups of pulsars which bridge sub-populations as well as pulsars which form new bridges across sub-populations requires efficient and sensitive pulsar and single-pulse searches, as well as regular monitoring of a large number of sources and joint studies of their emission properties at radio frequencies and beyond.

The idea of a complete census of Galactic NS with the SKAO was initially presented in the first SKA Science Book \citep{Cordes2004}.
A lot of the background and ideas discussed in this paper were further developed and presented in \cite{Tauris2015SKA} as part of the 2015 SKA Science book. Since then, the field has grown and evolved significantly. Here, we provide an update to the 2015 chapter, taking into account advances in the field, as well as changes made to the design of the SKAO telescopes in the last 10 years.

\section{Magnetars and High B-field Radio Pulsars}
\label{sec:magnetars}

Magnetars are NSs with extremely strong inferred surface magnetic field strengths. 
Their X-ray luminosity often exceeds the rotational energy budget, making their $10^{13}-10^{15}$\,G magnetic field their main energy resource \citep{DuncanThompson1992}.
They are frequently discovered as X-ray transients, which erupt with bursts, outbursts, and giant flares observed in X-ray and $\gamma$-ray emission. The energy required to produce such an outburst is higher than that available from spin down alone, and magnetars are instead thought to be powered by the decay of their enormous magnetic fields \citep{DuncanThompson1992}. 
There are currently about 30 NSs that showed magnetar-like behaviour (see the McGill Magnetar Catalog\footnote{\url{https://www.physics.mcgill.ca/~pulsar/magnetar/main.html} \citep{Olausen2014ApJS}} and \citealt{ReaDeGrandis2025}), 7 of which have also been detected in the radio band. The on-set of radio emission in these sources seems to be connected to their X-ray outbursts, and fade with time after the outburst. During the radio-active phase, the radio emission from magnetars is highly variable as observed in the flux density, linear and circular polarisation, pulse spin down, single-pulse emission, and spectral index. The variability is highest directly after the radio emission on-set, and stabilises with time until the radio emission seems to eventually turn off in most of the cases. 
 For recent reviews of magnetars and their emission, see e.g. \cite{Kaspi2017ARAA} and \cite{ReaDeGrandis2025}. 

The evolutionary connection between magnetars and other types of NSs is still a matter of debate. The evidence for magnetars not being a distinct class, but one sub-group of the larger pulsar family, is increasing. New sources are being discovered that exhibit magnetar-like behaviour, e.g. the CCO RCW103 \citep{Rea2016}, and high B-field radio pulsars like J1119-6127 \citep{Archibald2016} or the Kes75 PSR\,J1846-0258 \citep{Gavriil2008} , which have all shown magnetar-like outbursts in X-ray emission. The high B-field pulsars are sources that mostly appear as conventional rotational powered pulsars, but which have magnetar-strength inferred surface magnetic fields.  
Observations using X-ray telescopes, such as Chandra and XMM-Newton, have also shown that when comparing pulsars of similar age, the higher B-field ones show higher levels of thermal X-ray emission than lower B-field pulsars \citep{KaspiMcLaughlin2005, Zhu2011ApJ}. 

In the last few years, new magneto-thermal models have been developed, which aim to explain the apparent diversity of magnetars with other NS sources as an evolutionary pathway rather than a different type of object \citep[e.g.][]{Vigano2013, Gourgouliatos2016,Dehman2023b, Ascenzi2024,Igoshev2025}. This idea is also supported by the measurement of braking indices (defined as $n \equiv P \ddot{P} / \dot{P}^2$) for a few pulsars, e.g. PSR J1734--3333 \citep{Espinoza2011}, with a braking index of $n$\,=\,0.9 seems to be moving towards the magnetar region of the $P-\dot{P}$ diagram (see Fig. \ref{fig:ppdot_intro}). More measurements of this type will help to understand the connection between different classes of pulsars. If these new magneto-thermal models are correct, they will have a large impact on the understanding  of birth rates for both magnetars and for other sub-groups of NSs.   

Another important area of study is the connection of magnetars with other transient events, such as Gamma Ray Bursts, superluminous supernovae, and Fast Radio Bursts (FRBs). In particular, the possibility that magnetars generate at least some FRBs has been a major advance in the last few years. FRBs are extremely bright, millisecond duration radio bursts of extra-galactic origin \citep{Lorimer2007, Thornton2013}. Despite almost a thousand of these bursts having now been published, some repeating and some observed only once, the origin of FRBs is still unknown. One very promising theory involves single pulses of radio emission from magnetars, and with the discovery of an FRB-like single-pulse burst from the Galactic magnetar SGR 1935+2154 \citep{Bochenek2020, chimeFRB2020, CHIMEFRB-atel-2020}, it is increasingly likely that magnetars are the source of at least some FRBs. 

The high sensitivity of the SKAO telescopes will help address a range of questions relating to magnetars and high B-field pulsars, such as, are most magnetars truly radio quiet, and for those that have been observed in the radio band, has their radio emission really turned off in quiescence? Some studies have suggested that at least for some sources, this is really the case \citep[e.g.][]{bai2025}. 
Regular monitoring of magnetars and faint high-B pulsars for glitches, pulse profile changes, intermittency, timing variabilities, as well as more measurements of braking indices, would increase our understanding of the connection between the NS classes. 
New X-ray all-sky monitors \citep[e.g. the Einstein Probe,][]{Yuan2022} may soon identify many more high B-field pulsars via their X-ray outbursts. These sources will initially be characterised in the X-ray band by current and future X-ray telescopes, such as NewAthena \citep{Cruise2025}, and follow-up in the radio band will be crucial to form a complete picture of these sources. On the other hand, finding more high B-field pulsars with the SKAO telescopes will enable more follow-up X-ray observations, and this synergy with X-ray telescopes will add to our general understanding of NS.

\section{Central Compact Objects and Neutron Stars in Supernova Remnants}
\label{sec:CCOs}

The Central Compact Objects (CCOs) are young isolated NSs in young ($\lesssim$10 kyr) supernova remnants (SNRs) without known pulsar wind nebulae (e.g., \citealt{Pavlov2004}). There are currently 10 CCOs and at least 4 candidates \citep{Ferrand2012}\footnote{http://snrcat.physics.umanitoba.ca/SNRtable.php}, for a review, see \citet{DeLuca2017}\footnote{http://www.iasf-milano.inaf.it/~deluca/cco/main.htm}. CCOs have thermal X-ray spectra and have not been detected so far at radio frequencies (e.g., \citealt{Lu2024}) or other wavelengths. 
Based on their X-ray spectrum alone, CCOs are difficult to distinguish from quiescent magnetars and one (former) CCO in SNR RCW 103 has shown distinct magnetar-like behaviour as well as a 6.7-hr X-ray periodicity (e.g., \citealt{Rea2016}). 
Spin characteristics ($P \sim 0.1-0.4$\,s, $\dot{P} \sim 10^{-17}$\,s/s) have been measured for three CCOs (e.g., \citealt{Gotthelf2013,Halpern2010}) placing them in an underpopulated part of the $P - \dot{P}$ diagram (see Figure~\ref{fig:PPdot_SKA}) and implying surprisingly low dipole magnetic field strengths of $\sim 10^{10}$\,G.  
For two CCOs, \citet{Gotthelf2024, Gotthelf2020} the reported timing behaviour can be well modelled by one or more small glitches, or by extraordinary timing noise. 
CCOs are among the least understood young NS populations with different models proposing buried and re-emerging magnetic fields of normal or magnetar strengths, on-going accretion from fallback disks, or births with unusually low magnetic fields; e.g., \citet{Ho2011,Gencali2024,Halpern2010}. 

\citet{Gaensler2000} estimated the birth rate of CCOs to be at least $\sim 0.5$\,century$^{-1}$, marking CCOs as a significant contributor to the so-called ``NS birthrate problem'' \citep{Keane2008} -- that is that there are too many NSs in comparison to the known Galactic supernova rate if one assumes all the various NS manifestations are evolutionarily independent. 
To probe the CCO evolution and their unusual location in the $P - \dot{P}$ diagram, there have been a few unsuccessful searches for CCO descendants (so-called orphaned CCOs) \citep{Gotthelf2013b, Luo2015}. 
The best candidate for a CCO descendant is Calvera, a $P=59$\,ms NS with thermal X-ray emission, a supernova remnant association, and a magnetic field strength of  $4.4 \times 10^{11}$\,G that is intermediate between CCOs and the normal pulsar population \citep{Rigoselli2024}. So far, Calvera remains undetected at radio frequencies. 

The SKAO sensitivity and spatial resolution will allow understanding of CCOs and their links with the overall NS population in multiple ways:
\emph{(i)}   
The SKAO can determine whether the known CCOs are radio emitters and, if so, measure their spin periods and rate of spin-down. Sampling different CCO ages (as determined from the SNR ages) could then help probe the time of the onset of radio emission, in the cases where it was detected. For such studies more CCOs are required since the current CCO age span, $0.3 - 10$\,kyr, hinges on individual objects. 
\emph{(ii)}   
The SKAO can search for radio-emitting pulsars in SNRs, some of which may turn out to be CCOs. Timing of the newly discovered radio pulsars will help to constrain the birth period distribution of NSs.
There are $\gtrsim$300 known SNRs and $\gtrsim$200 SNR candidates \citep[e.g.][]{Anderson2017, Hurley-Walker2019, Dokara2021, Green2022}, a sample that keeps growing with the search of radio images using more sensitive radio telescopes. 
Recent deep MeerKAT searches (down to flux limits of $\sim$30 $\mu$Jy) of SNRs and CCOs illustrate the feasibility and potential of such deep radio surveys of SNRs (e.g., \citealt{Turner2024}). 
\emph{(iii)}   
The SKAO can also search for CCO descendants. This could be achieved through deep radio searches of enigmatic objects such as Calvera. A periodic and single pulse survey with the SKAO telescopes will also probe the underpopulated CCO location in the $P - \dot{P}$ diagram as well as the adjacent regions where CCO descendants may stage their radio appearance.
\emph{(iv)}   
The SKAO will be crucial in identifying and characterising new NS candidates found at other wavelengths such as X-rays. 
Multiwavelength synergy effects will be particularly important for CCOs considering the aforementioned possible links of CCOs with magnetars and the propensity to high-energy outbursts of the latter.

\section{Rotating Radio Transients and Intermittent Pulsars}

\begin{figure}[ht]
\centering
 \includegraphics[width=1.0\linewidth]{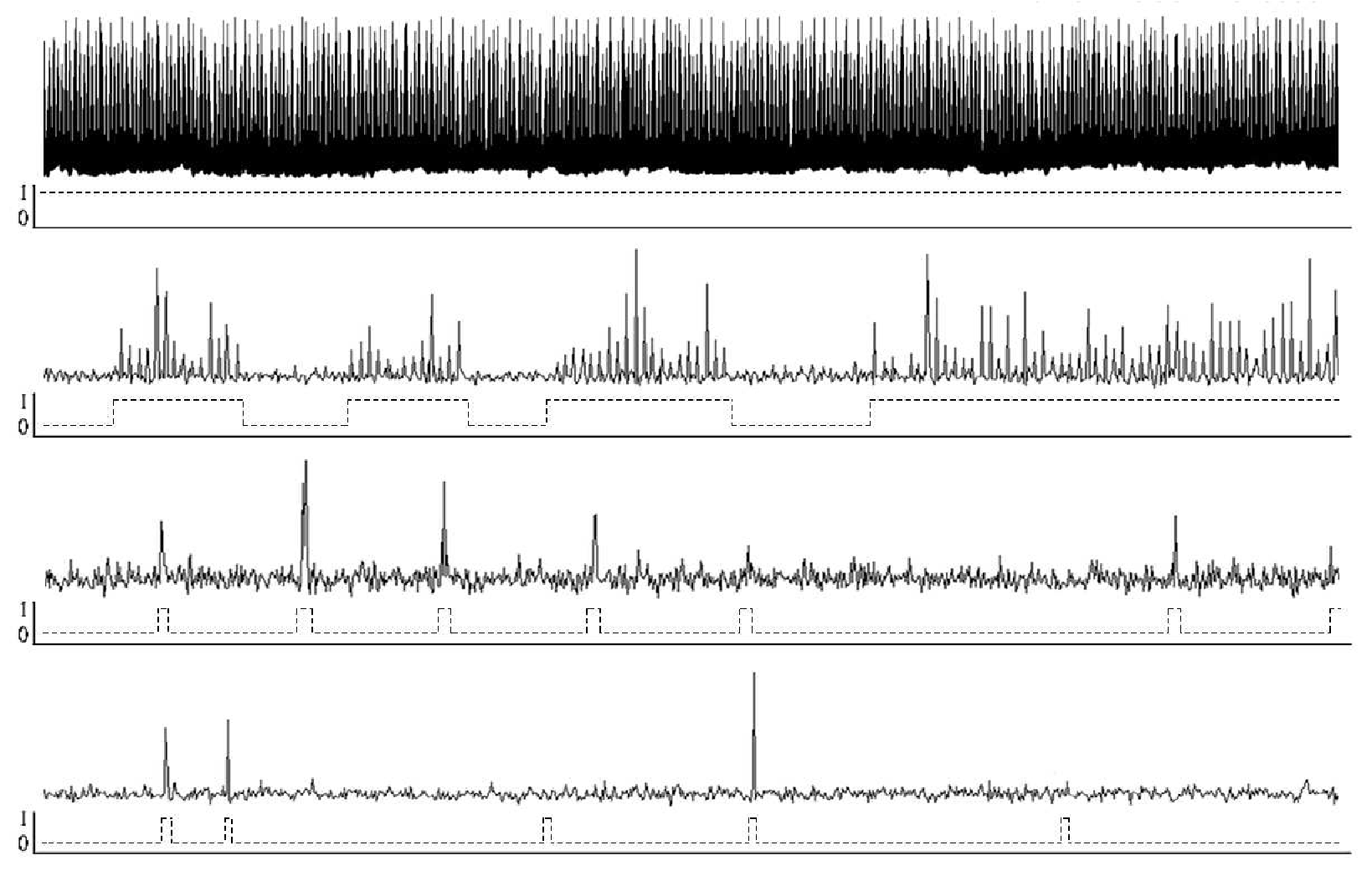}
 \caption{Time series (all of equal duration) taken from \citet{Burke-Spolaor2013} showing radio emission from a variety of sources 
(top to bottom: the Vela pulsar, PSR J1646--6831 (a nulling pulsar), RRAT J1647--36 and RRAT J1226--32). The
binary scales show an estimated representation of the null/emission state.}
 \label{fig:nullmodel}
\end{figure}

Since their discovery by \citet{McLaughlin2006} using single-pulse search techniques, rotating radio transients (RRATs) represent a complementary means of discovering and studying the spin-powered population of NSs. Some caution should be exercised since the simplest observational definition of an RRAT is sensitivity-dependent: for a given radio telescope, an RRAT is a pulsar
that is more easily detectable through a single-pulse search compared to traditional techniques designed to reveal periodic sources (most commonly Fast Fourier Transforms or Fast Folding Algorithms). It is possible to quantify this for both nulling and non-nulling pulsar cases \citep[as given in Equations 1 and 2 of][]{Burke-Spolaor2013}. A comparison of RRAT pulses versus pulses from other types of NSs is given in Fig.~\ref{fig:nullmodel}.

In the twenty years that have elapsed since the discovery of the RRATs,
single-pulse search algorithms are now an integral part of pulsar surveys. To date, while 211 RRATs are currently known\footnote{An up-to-date list of RRATs can be found within the ATNF pulsar catalogue.}, due to the difficulties in detecting them, not all of the parameters are well measured. In a recent statistical analysis, \citet{Abishek2022} found that, for the 42 RRATs with measured period derivatives ($\dot{P}$), these sources are preferentially found toward the upper right of the $B-P$ diagram, with an
average characteristic magnetic field strength of $6.3 \times 10^{12}$~Gauss
\citep[see also][]{Cui2017}. 
A recent population analysis
by \cite{Agarwal2025} attempted to constrain the underlying parameters of the population, which has many similarities with that of radio pulsars, although there is evidence for a larger scale height for RRATs: $600 \pm 200$~pc,  almost double that of the canonical pulsar population \citep[see, e.g.,][]{FaucherGiguere2006}.

The higher than average $B$ and $P$ values for the RRATs compared to the normal pulsar population is significant. The longest period 
reported for an RRAT so far is 11.9~s \citep{Zhou2023}, although there are known candidates with even larger periods (see below). An outstanding question is the connection between RRATs and other long-period pulsars, and the emerging population of long-period transients (see Section~\ref{sec:ULPs}). In addition, despite the small sample size currently observed, the Galactic population of RRATs is expected to be similar to that of normal pulsars \citep[see, e.g.,][]{McLaughlin2006,Agarwal2025}. A larger sample size as a result of SKAO surveys will not only constrain the populations, but also provide individual objects to study at other wavelengths. Currently, only
RRAT J1819--1458 has detectable X-ray emission \citep{McLaughlin2007}. The superior sensitivity of the SKAO, and other emerging instruments, is expected to play a role in a number of important, and currently poorly understood, aspects of the RRAT population. Firstly, in concert with the increasing sample sizes of other NS populations, enhanced statistics of the currently known RRATs, and the many that are expected to be discovered, should be able to shed light on the connection between pulsars, RRATs and long-period pulsars. We note in this regard that a recent MeerKAT survey by \citet{Turner2025} discovered an additional 26 RRATs and RRAT candidates, including RRAT candidate PSR~J2218+2902 which has a spin period of 17.5~s. For currently known RRATs, to improve constraints on the emission process(es), high-sensitivity observations are urgently required to probe the presence of emission during epochs where no pulses are currently detectable. We note that a combination of the precise localisation of the RRAT sources through real-time detection and imaging of single pulses and the commensal observing modes of the SKAO telescopes will greatly facilitate the measurement of the crucial $\dot{P}$ values, as demonstrated for SKAO precursors such as MeerKAT \citep{Turner2025}. A more complete census of the RRAT population could also elucidate the spectral properties of RRATs, provide braking indices through long-term timing, and the precise localisations will allow for rapid follow up at non-radio wavelengths.  

In addition to RRATs, a very important phenomenon in the pulsar population is that of pulsar intermittency. This behaviour, in which the pulsar switches between an ON and OFF state accompanied by an increase in spin-down rate during the ON-state, was first seen in PSR~B1931+24 by \citet{Kramer2006} and has subsequently been observed in only a handful of other pulsars \citep{Lorimer2012,Camilo2012,Lyne2017}.  
A larger number of pulsars show switching between the ON and OFF states at shorter time scales, from seconds to hours, and are generally called nulling pulsars. These state changes are too short to allow us to see any potentially associated changes in the spin-down rate.
Some pulsars also show switching between different modes of emission, which manifest as differences in pulse profiles or flux density. Such sources are called mode-changing pulsars, and some of the ones with the longest mode-changing timescales have also shown different spin-down rates for different modes \citep[e.g.][]{Lyne2010}. 
Since these different groups of sources have such similar properties, it is easy to suspect that the emission changes have the same underlying origin. 
Many models have been developed to explain the phenomena; some consider the plasma in the magnetosphere and how it is connected to the global charge distribution \citep[e.g.][]{Timokhin2010, Kalapotharakos2012, Li2012, Melrose2014}, others suggest variations in the particle acceleration region \citep{Szary2015} or a twisted magnetosphere \citep{Huang2016}. 

Intermittent, mode-changing, and to some extent nulling pulsars, provide excellent opportunities to study how a pulsar's spin-down is related to its radio emission. It is therefore very important to find more of these types of sources. The SKAO telescopes, with their excellent sensitivity and wide field of view, will be significant in this search. In addition, high-sensitivity observations with the SKAO telescopes will be able put better limits on the radio emission in the OFF-state, and by using the multibeaming capabilities of the pulsar timing backend, higher cadence observations will help in determining more exact switch times between states in known intermittent pulsars.

\section{Isolated Pulsar Evolution on the P-Pdot Diagram}
\label{sec:PPdot}

\begin{figure*}[t!]
\centering
\includegraphics[width=0.42\linewidth]{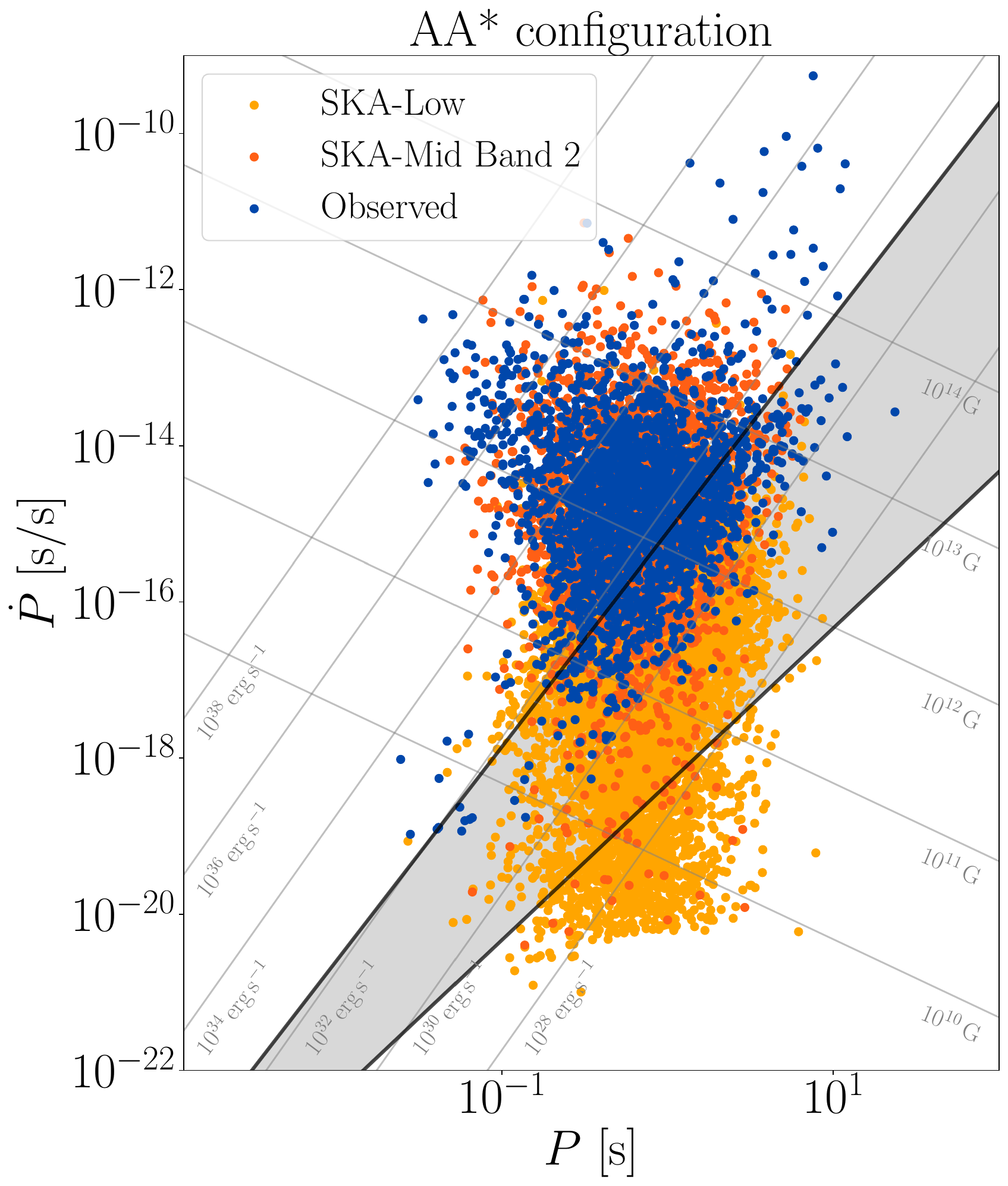}
\hspace{0.2cm}
\includegraphics[width=0.42\linewidth]{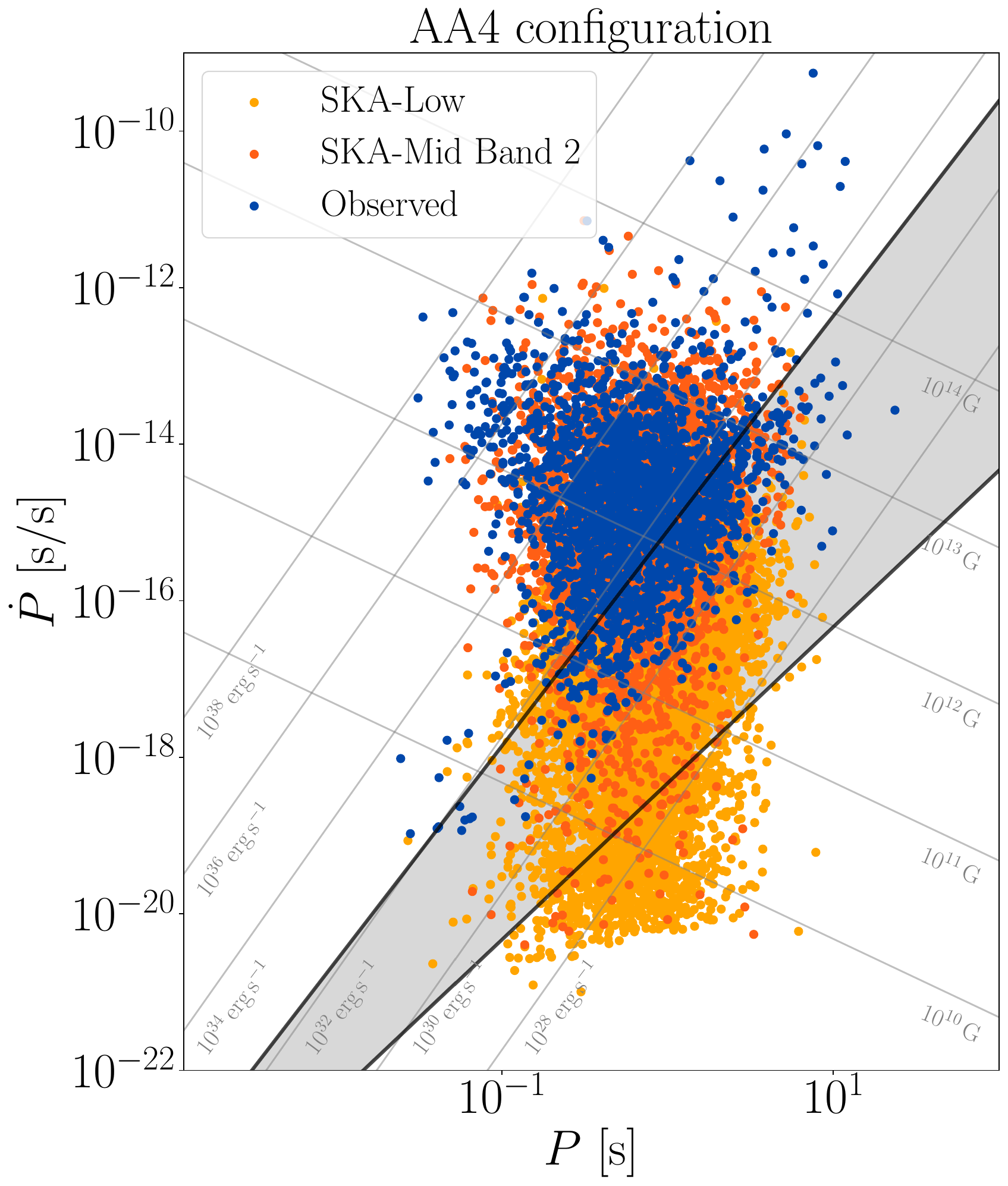}
\caption{$P$-$\dot{P}$ diagrams for the expected population of isolated pulsars observed with SKAO in the AA* (left) and AA4 (right) configurations corresponding to Survey Option 3 for the evolutionary population synthesis approach outlined in~\citet{Keane2025_SKA_Census}. We show the currently observed population in blue (data taken from the ATNF Pulsar Catalogue v2.5.1; \citet[][]{Manchester2005}, \url{https://www.atnf.csiro.au/research/pulsar/psrcat/}), and pulsars detected with SKAO Low and Mid Band 2 in yellow and orange, respectively. Lines of constant rotational energy loss and dipole magnetic field strength are shown in grey. The black lines show extreme versions of the pulsar death lines below which radio emission has been proposed to cease and delimit the so-called ``death valley'' shaded in grey \citep{Chen1993}. Note that these population synthesis simulations take into account the natural fading of the radio emission but do not account for sudden emission switch-off or occasional switch-on of stronger emission that may be detectable with single-pulse searches (for a recent example, see \citealt{Rajwade2025arXiv}).}  
\label{fig:PPdot_SKA}
\end{figure*}

Observing pulsars provides a snapshot of a fraction of the pulsar population at present. As NSs have formed throughout the Galaxy's history, any such snapshot provides information about NSs across a range of ages. Studying present-day properties for a sufficiently large number of stars, thus allows us to constrain population-level properties. 
The increase in the known population as resulting from planned pulsar surveys with the SKAO telescopes will hence play a crucial role in our understanding of the population as a whole. In this section, we will summarise the current state of evolutionary NS population modelling and present predictions for the isolated pulsar population observable with the SKAO in the AA* and AA4 configurations.

A key diagnostic for analysing pulsar populations is the period-period derivative ($P$-$\dot{P}$) diagram. This is particularly crucial when focusing on isolated pulsars, as the $P$-$\dot{P}$ diagram enables the identification of distinct NS classes and possible evolutionary relationships. Mapping between the observed population (currently a few thousand isolated sources predominantly seen in the radio) and the underlying population also requires the incorporation of observational biases and limitations~\citep{Keane2025_SKA_Census}. Once incorporated, these population synthesis approaches allow us to extract information on NS birth properties, birth rates, and the fundamental laws that govern their evolution \citep{Lorimer2004, FaucherGiguere2006,Bates2014,Johnston2017}. Recent advances in this area have involved the implementation of updated theoretical models, the increase of available computational resources, and adaption of machine learning techniques.

For isolated NSs, birth parameters shaping the $P$-$\dot{P}$ distribution fall into two categories: dynamical properties and magneto-rotational characteristics. The birth locations of pulsars are relatively well understood \citep{Yusifov2004}, but uncertainties remain regarding the kick velocity distribution imparted during the supernovae (see Section~\ref{sec:NSkicks_vel} below). This will primarily affect future all-sky surveys which will detect older pulsars far from the Galactic plane. Magneto-rotational properties, in turn, depend on natal magnetic field strengths, $B$, and initial rotation periods. The former are typically assumed to follow a log-normal distribution \citep{FaucherGiguere2006, Popov2010,Gullon2014, Gullon2015, Cieslar2020, Igoshev2022B}, with recent studies inferring $\mu_{\rm log \, B} \sim 13.1$ and $\sigma_{\rm log \,  B} \sim 0.5$ \citep{Graber2024, PardoAraujo2025} based on updated magnetic-field evolution models (see below). 
The functional form of the period distribution, initially assumed to be Gaussian \citep{FaucherGiguere2006, Cieslar2020}, is more difficult to constrain. This is due to the coupled evolution of spin period, magnetic field strength and misalignment angle and the fact that NSs lose memory of their initial spin periods after a characteristic evolutionary time-scale~\citep[see][for details]{Graber2024}. Based on simplifying assumptions of the pulsar spin-down, two recent analyses of around 70 young pulsars in supernova remnants found evidence for a log-normal \citep{Igoshev2022B} and a Weibull distribution \citep{Du2024}, with differences arising from the treatment of the underlying observational selection effects. Based on the log-normal $P$ distribution, two neural network-based studies have inferred $\mu_{\rm log \, P} \sim -1$ to $-0.7$ and $\sigma_{\rm log \, P} \sim 0.4$ to $0.6$ \citep{Graber2024, PardoAraujo2025}.

While dynamical properties are easily evolved via the solution of Newtonian equations of motion in the Galactic potential, different approaches have been used to evolve magneto-rotational properties. Many earlier works assume the NS spin-down to be dipolar, neglect misalignment angle variations, and do not account for the possibility of magnetic field decay. While such assumptions enable fast generation of synthetic pulsar populations, they neglect key physics of realistic NSs. In particular, simulations of plasma-filled magnetospheres \citep{Spitkovsky2006, Philippov2014} indicate that the spin-down is governed by two coupled differential equations, combining spin-down contributions from the magnetic dipole and the plasma currents,
\begin{equation}
	\dot{P}(t) = \frac{\pi^2}{c^3}\frac{B(t)^2 R^6}{I P(t)} \left[1 + \sin^2 \chi(t) \right],
\end{equation}

\begin{equation}
    \dot{\chi}(t) = -\frac{\pi^2}{c^3}\frac{B(t)^2 R^6}{I P(t)^2} \, \sin\chi(t) \cos\chi(t),
\end{equation}
where the misalignment angle $\chi$ is distributed uniformly at birth. Here, $c$ denotes the speed of light, and $R$ and $I$ the NS radius and moment of inertia, respectively. Assuming these equations are correct, they imply that $\chi$ decreases as pulsars slow down (for other models, see \citealt{Novoselov2020,Toropov2025}). The exact behaviour of such evolutionary models, however, also requires a prescription for magnetic field evolution, $B(t)$, commonly assumed to be exponential or power-law like \citep{Aguilera2008}, based on Ohmic dissipation and the Hall effect, respectively. While these provide general insights into field changes, multi-dimensional simulations are ultimately needed to accurately capture field evolution in NSs \citep{DeGrandis2020, Vigano2013, Vigano2021, Dehman2023a, Ascenzi2024}. Several recent studies incorporate two-dimensional magneto-thermal simulations into pulsar population synthesis, finding field decay crucial to match observations~\citep[e.g.,][]{Cieslar2020, Gullon2014, Dirson2022, Graber2024, Shi2024}. While early-time field evolution is less important for radio pulsars, it is essential for modelling young magnetars as suggested by \citet{Gullon2015, Sautron2025}. Including magnetars and radio-quiet NSs in future population synthesis pipelines will thus help improve natal field constraints for strongly magnetised sources, poorly constrained by radio observations alone.

Accounting for misalignment angle and magnetic field evolution, pulsars are expected to migrate from the top-left to the bottom-right in the $P$-$\dot{P}$ plane as they age. 
As they do so, the stars' available spin-down power decreases. 
In addition, narrower pulses are measured with increasing pulse periods for most pulsars \citep{Posselt2021}. This can be interpreted as shrinking emission beams.
Consequently, several modern population synthesis approaches can reproduce the observed radio pulsar population without invoking a ``death line'' where radio emission ceases \citep{Gullon2015, Graber2024, PardoAraujo2025, Shi2024}. However, the fading of old NSs and their $P$-$\dot{P}$ locations also depend on radio emission models. The radio luminosity, $L$, is typically assumed to be proportional to the spin-down power, $L \propto |\dot{E}_{\rm rot}|^{\alpha} \propto P^{-3\alpha} \dot{P}^{\alpha}$ or more generally $\propto P^{\beta} \dot{P}^{\gamma}$. For example, \citet{Cieslar2020} found $\beta \sim -0.37$ and $\gamma \sim 0.26$, while \citet{PardoAraujo2025} inferred $\alpha \sim 0.7$ \citep[see also][]{Shi2024} and \citet{Posselt2023} obtained $\alpha \sim 0.15$. These estimates are, however, not directly comparable, as \citet{Cieslar2020} model the pseudo-luminosity and introduce a period-dependent beaming factor which is set as a constant by \citet{Posselt2023}, while \citet{PardoAraujo2025} considers the intrinsic luminosity, a time-dependent beaming model and pulse propagation through the interstellar medium.  Figure~\ref{fig:PPdot_SKA} summarises our current understanding of isolated pulsar evolution and makes predictions for SKAO observations with the AA* and AA4 baselines (see Section \ref{sec:obsmodes}). We highlight that the SKAO telescopes will detect many faint, old pulsars in the lower-right region of the $P$-$\dot{P}$ plane, enabling us to refine highly uncertain emission models and death-line physics~\citep{Oswald2025_MAG}. This is especially relevant for the newly discovered class of long-period radio emitters (see Section~\ref{sec:ULPs}). Consistent radio flux measurements alongside period and period derivative data will be crucial to understanding their emission \citep{Cieslar2020, PardoAraujo2025, Keane2025_SKA_Census}. Moreover, these low $\dot{P}$ sources will also be interesting for continuous gravitational wave searches and exploring the `spin-down limits' of the corresponding gravitational wave strain~\citep{Abac2025}.

While population synthesis broadly captures the observed radio pulsar population, not all pulsars move downward in the $P$-$\dot{P}$ plane. Measuring braking indices, 
reveals deviations from the dipolar spin-down expectation of $n = 3$. While some of these are readily explained by field decay, several young pulsars show $n \simeq 2$ \citep{Espinoza2017} or well below \citep[e.g.][]{Espinoza2011}, indicating upward motion in the diagram. This may be linked to magnetic field growth \citep{Ho2015} but is complicated by pulsar glitches affecting long-term spin evolution \citep{Lower2021} and general long-term variability \citep{Lower2025}. Although such young sources may provide insights into internal field evolution and dense matter physics~\citep{Basu2025_SKA_EOS}, their impact on future pulsar population studies is likely minimal, as the SKAO telescopes will predominantly probe older sources that are less prone to exhibit glitches. However, such effects may be relevant when modelling central compact objects, magnetars, and high-B field pulsars (see Sections~\ref{sec:magnetars}--\ref{sec:CCOs}). In general, monitoring of time-variable braking indices for the bulk of the pulsar population may shed light on the evolution of these sources within the $P$-$\dot{P}$ plane and the validity of modelling assumptions such as inferred magnetic field strengths and characteristic ages.

As isolated pulsar modelling grows in complexity, robust inferences become more challenging. Traditional statistical techniques such as Kolmogorov-Smirnov tests, $\chi^2$ analyses, and Markov Chain Monte Carlo (MCMC) methods struggle with high-dimensional parameter spaces and complex simulation frameworks. Recent studies \citep{Graber2024, PardoAraujo2025, Sautron2025}, highlight the promise of simulation-based inference with neural networks for overcoming these challenges. In the future, machine-learning approaches may also finally enable robust comparisons between physical models, which has not been possible to date. An SKAO telescope pulsar census (see Section~\ref{sec:obsmodes} and ~\citealt{Keane2025_SKA_Census} for details) will be instrumental in advancing these studies by providing comprehensive, consistent pulsar measurements, particularly when combined with all-sky surveys in other wavebands.

\section{Pulsar Recycling and Formation of Exotic Binaries and Triples}
\label{sec:binaries}

\begin{figure}[ht]
 \includegraphics[width=1.0\linewidth]{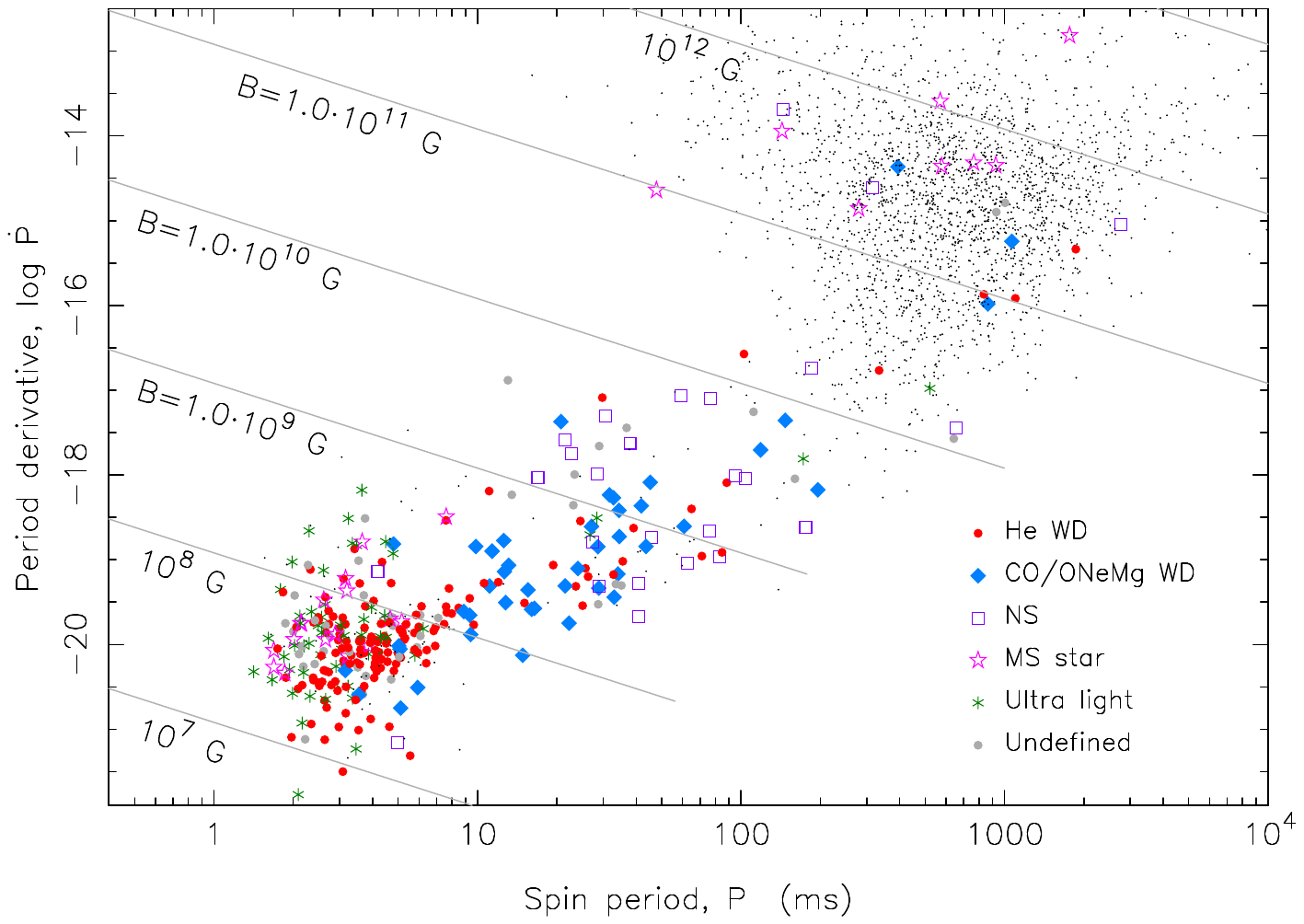}
 \caption{Distribution of 346 binary radio pulsars in the $P\dot{P}$-diagram.
    The nature of the companion stars is indicated with different symbols. Isolated pulsars are represented with a dot. Lines of constant surface B-field flux density are shown. Data taken from the {\em ATNF Pulsar Catalogue} version 2.6.0 in February~2025
    \citep[][\url{https://www.atnf.csiro.au/research/pulsar/psrcat}]{Manchester2005}.}
 \label{fig:pulsar-companions}
\end{figure}

The widely accepted formation channel for fast-spinning millisecond pulsars (MSPs) is the recycling scenario \citep{Alpar1982,rs82,Bhattacharya1991}, in which a NS is spun up through mass and angular momentum accretion from a companion star. The outcome depends on the initial binary configuration and subsequent evolutionary phases \citep[see][for a detailed review]{tv23}, leading to recycled pulsars with spin periods ranging from just a few milliseconds ('fully recycled') to several tens or hundreds of milliseconds ('mildly recycled'). These NSs can orbit a diverse set of companion stars (see Figure~\ref{fig:pulsar-companions}) --- ranging from compact objects (white dwarfs, NSs, or black holes) to main-sequence stars, or even ultra-light planetary-mass bodies. Observationally, of the 638 recycled pulsars discovered with $P<30\;{\rm ms}$ \citep{Manchester2005} --- 426 in the Galactic disk and 212 in globular clusters (GCs) --- nearly 60\% are found in binary systems, most commonly with helium white dwarf (He-WD) companions in circular orbits.
For comparison, the binary fraction among young radio pulsars is estimated to be at most 5--8\% \citep{Antoniadis2021}.

A small but growing number of MSPs with unusual properties have been discovered, including the 'triple pulsar' PSR~J0337+1715, which is orbited by two WDs \citep{Ransom2014,vcs+25}; PSR~J1903+0327, which has a main-sequence companion \citep{fbw+11}; the now established class of enigmatic  eccentric MSPs (eMSPs), potentially indicating a hierarchical triple system origin
\citep{Champion2008,dsm+13,Antoniadis2016,Barr2017,Serylak2022,Grunthal2024}.
Another intriguing case is the MeerKAT discovery of PSR~J0514$-$4002E (Barr, Dutta, et al.\ \citeyear{bd+24}), which orbits either a low-mass BH or a very massive NS --- likely the remnant of a former merger event in the dense environment of its host GC.

Thanks in large part to the launch of the Fermi $\gamma$-ray Space Telescope --- which has provided valuable targets for deep follow-up pulsar searches in the radio band \citep{Fermi2009} --- an ever-growing number of MSPs has been discovered in binaries with light, semi-degenerate companions. These pulsars are often characterised by eclipses of their pulsed radio signal over large portions of the orbit, caused by matter surrounding the system. This material is either ablated from the companion star by the pulsar wind or may represent residual mass still being ejected from the companion’s Roche lobe. These so-called spiders \citep{rob13} are classified into two subgroups: {\em black widows}, with companion masses typically well below 0.1\,M$_\odot$, and {\em redbacks}, with companion masses of $\sim$\,0.1–0.3\,M$_\odot$.
It remains debated whether these are two distinct populations \citep{ccth13}, or instead represent an evolutionary sequence \citep{bdh14,mly25}. 

A compelling evolutionary link between accretion-powered and rotation-powered binary pulsars is provided by the discovery of the so-called 'transitional MSPs' (tMSPs), which alternate between accretion-powered high-energy pulsed emission and rotation-powered radio pulsations \citep{Archibald2009Sci,pfb+13}. These systems offer the most direct and compelling observational evidence in support of the recycling scenario.
While there is now ample evidence that this scenario is broadly correct, many details remain poorly understood. Some of the most significant open questions \citep[e.g.][and references therein]{tlk12} include: the mechanisms behind accretion-induced decay of the surface magnetic field \citep[but see also][]{Cruces2019}; the maximum attainable spin rate of an MSP; the Roche-lobe decoupling phase; the nature and physics of accretion torque reversals; and the spin-up line.

\begin{figure}[ht]
\centering
\includegraphics[width=1.0\linewidth]{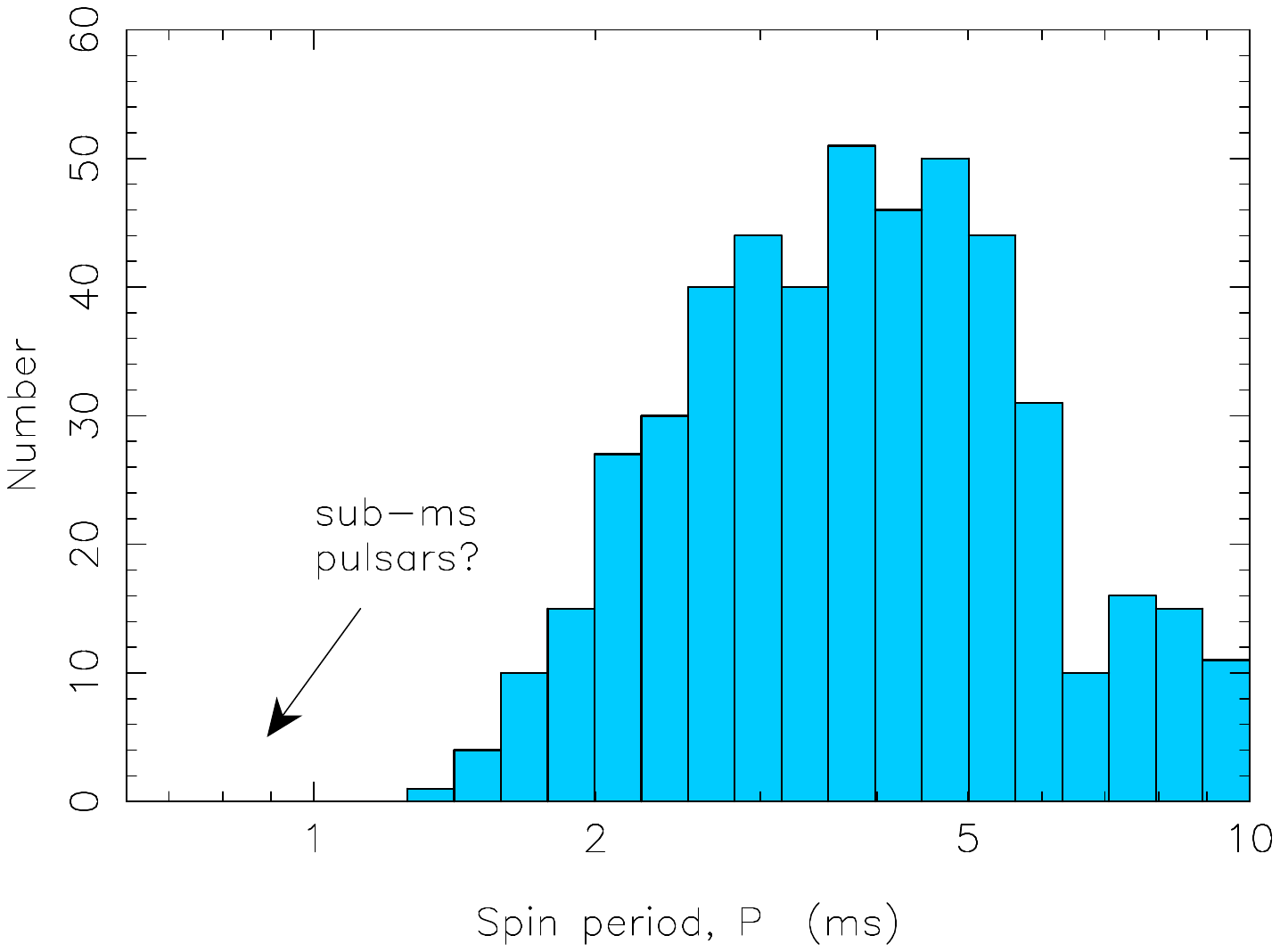}
 \caption{Distribution of spins of 485 radio MSPs with $P < 10\;{\rm ms}$. After \citet{tv23}.}
 \label{fig:MSP-spins}
\end{figure}

The SKAO-Mid telescope will be particularly well suited for detecting ultra-compact binary pulsars in the final stages of their recycling --- potentially revealing the long-sought sub-ms pulsar (see Figure~\ref{fig:MSP-spins}). The MSP currently known to have the fastest spin rate PSR~J1748$-$2446ad, has a period of 1.4~ms \citep{hrs+06}. Its location in a globular cluster and its extreme spin raise the question of whether multiple recycling episodes in a dense stellar environment \citep{vf14} might be a necessary condition for producing a sub-ms pulsar.

Over the past decade, the number of known recycled pulsars has nearly doubled. A key driver of this growth has been the improved sensitivity of radio instrumentation. Historic telescopes --- such as the Murriyang telescope at Parkes in Australia and the Effelsberg telescope in Bonn, Germany --- have been upgraded with ultra-wideband receivers, while new, highly sensitive facilities like the FAST telescope in China and the MeerKAT array in South Africa have significantly increased the discovery rate. These advances have also enabled the detection and detailed study of increasingly peculiar systems.

Improvements in instrumentation often lead to significantly higher data rates, driven by larger bandwidths combined with higher time and frequency resolution. As a result, the development of reliable and efficient software for processing the growing volumes of pulsar data has been a crucial factor in the continued success of pulsar science --- especially for the discovery of MSPs in compact binaries, which requires searching over a vast parameter space. Key advances include acceleration and jerk search techniques, GPU-accelerated pipelines such as PRESTO \citep{ran11}, and template banking algorithms \citep[e.g. {\em Peasoup},][]{bar20}.

\section{Double Neutron Star and Neutron Star-Black Hole Systems}

Double neutron star (DNS) systems represent the rare and extreme endpoint of an evolutionary journey involving two massive stars and their binary interactions. Currently, more than two dozen DNS systems are known (including unconfirmed candidates), but only in one --- {\em the double pulsar} J0737$-$3037 \citep{bdp+03,Lyne2004} --- has radio pulses been observed from both NSs. 
Many additional DNS systems are expected to be discovered with the SKAO telescopes in upcoming pulsar surveys \citep{kbk+15, Keane2025_SKA_Census}. In combination with the LIGO–Virgo–KAGRA detections of DNS mergers, beginning with GW170817 \citep{2017PhRvL.119p1101A} and GW190425 \citep{2020ApJ...892L...3A}, the coming decade offers a unique opportunity to refine our understanding of the formation and evolution of DNS systems \citep{tkf+17}. This includes their accretion history during the high-mass X-ray binary stage, the common-envelope and spiral-in phase, and the subsequent so called Case BB mass transfer phase \citep[see e.g.][]{Delgado1981}, in which the first-born NS is mildly recycled while stripping the envelope of its He-star companion prior to core collapse and SN explosion \citep{Tauris2015MNRAS}.

Measurements of DNS masses, spins, orbital characteristics, and velocities provide crucial insights into their formation history and the nature of NS kicks, and are also essential for testing fundamental theories of gravity \citep{Kramer2021}.
The MeerKAT discovery of PSRJ0514$-$4002E (Barr, Dutta, et al.\ \citeyear{bd+24}), orbiting either a low-mass BH or a massive NS, represents the closest realisation to date of detecting an MSP in orbit around a BH. This source is located in the GC NGC~1851, supporting a formation scenario involving an exchange encounter.
Despite population synthesis predictions \citep[e.g.][]{csh2021}, such MSP+BH systems are not expected to form frequently in the Galactic field due to the short nuclear timescale of the BH progenitor, which limits the duration and efficiency of the X-ray recycling phase. Therefore, continued targeted searches in GCs or in the Galactic Centre, where similarly large stellar densities can be found, appear to offer the best chance of detecting a compact MSP+BH binary~\citep[e.g.][]{Liu2014,Bagchi2025_SKA_GlobClust, Abbate2025_SKA_GalCen} -- unless such a system is ejected into the field, which remains a possibility.

Over the past decade, gravitational wave (GW) observations have finally revealed mergers involving extragalactic DNS and NS+BH systems, and have unequivocally linked them to short $\gamma$-ray bursts \citep{sfk+17,aaa+18c}. The observed properties of these merger events, together with their likely connections to Galactic DNS systems, highlight the need to better constrain the properties and formation rates of Galactic DNS and NS+BH binaries. Improved understanding of these populations will refine theoretical predictions of GW merger rates \citep[e.g.][]{bkr+08,ktl+18}, enabling more robust comparisons with observations and providing critical feedback on population synthesis models.
In addition, the past decade has brought significant advances in strong-field tests of gravity using DNS systems \citep{Kramer2021}. On the theoretical side, the first detailed simulations of close binary stellar evolution up to the onset of core collapse --- leading to the second, ultra-stripped SN and the final formation step of a DNS system --- have been carried out using MESA \citep{jtcf21}.

A critical challenge in discovering pulsars in tight, eccentric binaries (such as the DNS, MSP+BH, and NS+BH systems) is their detectability: such binaries are difficult to find without enhanced acceleration or acceleration–jerk searches \citep[e.g.][]{blw2013, ar2018}, or alternative methods such as template bank algorithms that account for all five Keplerian parameters \citep{bcb2022}.
Additional complications may arise from relativistic spin-orbit coupling \cite[e.g.][]{Stairs2004} or perhaps even light-deflection \citep{Doroshenko1995,Rafikov2006,Hu2022,db2023}, which can alter the pulse profile across the orbit, thereby complicating the data analysis.
However, the enhanced sensitivity of the SKAO telescopes, combined with the planned pulsar acceleration searches, will allow for high signal-to-noise pulse profiles, helping to overcome these obstacles.

\section{Neutron Star Masses}

High-precision radio pulsar timing is currently the most effective technique for measuring NS masses. By tracking every rotation of a pulsar over months to decades, timing models allow us to extract orbital information and, in few favourable binary systems, relativistic post-Keplerian parameters that are directly sensitive to the masses of the system’s components~\citep{Damour1992}. In particular, measurements of the Shapiro delay have now enabled mass estimates with uncertainties well below 15\% (see Figure 3 in~\citealt{Basu2025_SKA_EOS}).

Over the past 15 years, these timing observations have significantly expanded the NS mass distribution, confirming the existence of both high-mass and low-mass NSs. The most massive NS discovered through pulsar timing to date is PSR~J0740$+$6620 with a mass of $2.08 \pm 0.07\,M_\odot$~\citep{Fonseca:2021wxt}. Discoveries like this have set critical benchmarks on the maximum attainable NS mass, specifically ruling out those nuclear-matter equations of state that are particularly soft. Additional mass constraints have come from multi-wavelength observations in spider pulsar systems, as discussed above, which suggest even heavier NSs may exist~\citep{Linares:2019aua}. Currently, the heaviest spider pulsar is PSR~J0952$-$0607 with a mass of $2.35 \pm 0.17 M_{\odot}$~\citep{Romani2022}. 
However, to measure the mass of a spider pulsar, spectroscopy of the companion star is used to determine its radial velocity, which in turn can provide an inclination-dependent mass measurement. Estimates of the system's inclination can then be inferred from the optical light curve, but the resulting masses are often overestimated due to uncertainties in the  modelling of the heat distribution on the surfaces of the NS companions~\citep{Voisin2020b, Clark2023}.

At the low-mass end, systems like PSR~J0453$+$1559, the lightest NS candidate detected with pulsar timing with $1.174 \pm 0.004\,M_\odot$~\citep{Martinez2015}, challenge our understanding of stellar core collapse and binary evolution~\citep{Tauris2019, Muller2025} suggesting possible diversity in NS formation channels. Neutron star minimum masses below $1M_{\odot}$ as recently suggested by spectral modelling of the central compact object in a supernova remnant HESS~J1731$-$347~\citep{Doroshenko2022} would question our understanding of NS formation even further. However, these mass measurements are based on numerous assumptions and larger masses can explain
the data equally well~\citep{Alford2023}. Future radio timing observations with the SKAO will be crucial to narrow down the full NS mass distributions.

More recently, mass measurements have also been achieved by analysing gravitational waves from binary NS mergers. As the signal depends directly on the chirp mass, dependent in turn on the two component masses, a corresponding measurement together with assumptions on the NSs' spin, allows constraints on the two NS masses. For GW170817, corresponding mass estimates are in line with those obtained from radio pulsar timing, albeit less certain~\citep{2017PhRvL.119p1101A}. Gravitational wave observations have also opened up the study of objects in the so-called lower ``mass gap'' between $2-3~M_\odot$ (e.g., the secondary object in GW190814; \citealt{Abbott:2020ApJ}), where it is unclear whether the compact object is a NS or a black hole. Note that a similar mass-gap object, the companion of PSR~J0514$-$4002, has now also been detected via radio timing~\citep{bd+24}, highlighting that detections in similar mass ranges can also be expected in the SKAO era. Objects with these masses directly probe the maximum NS mass and corresponding formation scenarios of compact objects.

Beyond understanding the nuclear equation of state~\citep{Basu2025_SKA_EOS}, massive NS binaries also serve as laboratories for gravity~\citep{FreireWex2024}. Timing experiments in relativistic binaries have provided stringent tests of general relativity and alternatives, especially when multiple post-Keplerian parameters can be measured. The famous double pulsar system~\citep{Kramer2021}, along with pulsar--white dwarf binaries and triple systems~\citep{Voisin2020}, has yielded some of the most precise tests of gravitational theories to date.

Looking ahead, the SKAO is expected to revolutionise this field. With its unprecedented sensitivity and wide-field capabilities, SKAO will perform deep, all-sky pulsar surveys that are expected to greatly increase the number of known pulsar binaries~\citep{Keane2025_SKA_Census}. This will drastically expand the sample size of precisely timed pulsars, improving statistical analyses of the NS mass distribution and offering new opportunities for detecting extreme systems---either with exceptionally high or low masses. The discovery of new relativistic binaries through these surveys will thus enable new tests of general relativity and provide valuable data on the demographics of NS populations across different formation channels. Finally, the participation of SKAO in sensitive VLBI observations can provide the precise parallax measurements needed to calibrate distance-dependent contributions to timing observables used to test post-Newtonian gravitational theories \citep[e.g.,][]{Kramer2021}.

Moreover, synergies with other observatories will further enhance equation of state constraints. X-ray observations like those performed with NICER already provide complementary radius estimates via pulse profile modelling~\citep{Salmi24a} (especially powerful when combined with tight mass priors from pulsar timing), while gravitational wave detections from compact binary mergers contribute additional mass information. While the latter are currently uncertain, next-generation facilities will allow us to better probe the NS masses in the gravitational wave band~\citep{Abac:2025saz}. Together, these approaches will eventually allow us to map out the full NS mass spectrum, refine the maximum and minimum NS mass, and deepen our understanding of their interiors and their role in testing gravity under extreme conditions.

\section{Neutron Star Kicks and Velocities}
\label{sec:NSkicks_vel}

A few years after the discovery of radio pulsars, \citet{go70} noted that the vertical distribution of radio pulsars around the Galactic plane is much broader than that of their OB star progenitors, which are confined to within a few hundred pc of the plane. To explain this discrepancy, they introduced the concept of a 'velocity kick' --- a momentum impulse imparted to the newborn NS during the SN event.
In a seminal paper, \citet{Lyne1994Natur} convincingly showed that radio pulsars receive substantial natal kicks, with mean velocities around $\sim 450 \pm 90\;{\rm km\,s^{-1}}$, based on measurements of their peculiar motions relative to the local standard of rest. This was followed up by \cite{Hobbs2005MNRAS}, and more recently, among others, by \citet{Verbunt2017AA} and \citet{Igoshev2020MNRAS}.

The physical mechanism responsible for a pulsar’s natal kick remains under debate. Proposed explanations include asymmetric mass ejection driven by hydrodynamic instabilities during the SN explosion \citep{Janka2017ApJ}; anisotropic neutrino emission influenced by strong magnetic fields \citep{Chugai1984SvAL,Arras1999ApJ}; and the so-called 'electromagnetic rocket' effect resulting from an off-centre dipolar magnetic field \citep{Harrison1975ApJ,Agalianou2023MNRAS}.
In recent years, natal kicks with large amplitudes have been successfully reproduced in 3D SN simulations \citep[e.g.][]{Burrows2024} which, lately, also try to explain an observed correlation between the NS spin and velocity vectors
\cite[see e.g.][]{Johnston2007,Yao2021,Janka2022}.

Both theoretical and observational evidence suggests that natal kick amplitudes depend on the NS’s formation pathway. Theoretically, electron-capture SN models predict kicks of less than $10\;{\rm km\,s^{-1}}$ \citep{gj18}. In tight binaries, where an ultra-stripped SN forms the second-born NS, models indicate that the minimal mass loss involved often (but not always) results in weak natal kicks \citep{Tauris2015MNRAS}.
Observationally, small kicks are supported by the presence of large populations of radio pulsars in globular clusters,\footnote{\url{https://www3.mpifr-bonn.mpg.de/staff/pfreire/GCpsr.html}} which have escape velocities below $50\;{\rm km\,s^{-1}}$. Similarly, small NS kicks are invoked to explain the overabundance of $r$-process elements in ultra-faint dwarf galaxies, which have even lower escape velocities of $\sim 15\;{\rm km\,s^{-1}}$, as a result of binary NS mergers \citep[e.g.][and references therein]{Ding2024}. One of the best pieces of evidence for the existence of small kicks is the measured low transverse velocity of the double pulsar with a velocity of $11.1\pm1.0\;{\rm km\,s^{-1}}$ and the very small inclination angle between the pulsar spin and the total orbital angular momentum vector of less than 3.2 deg \citep{Kramer2021}.
Conversely, large kicks are required to explain young pulsars --- along with their associated bow shocks in SN remnants --- moving at velocities exceeding $1000\;{\rm km\,s^{-1}}$ \citep[e.g.][]{crl93,skr+19}.
While it is therefore likely that kick magnitudes are linked to the final structure of the progenitor star and the nature of the SN explosion, the detailed mechanisms remain uncertain.

The main technique for constraining the peculiar velocity --- and thus the natal kick --- of an individual radio pulsar involves combining a proper motion measurement with either a direct distance measurement (via geometric parallax) or a model-dependent distance estimate based on dispersion measure (DM). Proper motion and parallax can be obtained through either pulsar timing or interferometric imaging. 
For MSPs, proper motion can be measured to high precision even with a modest timing baseline. Non-recycled pulsars observed over many years can also yield meaningful proper motion constraints \citep{Hobbs2004MNRAS,Hobbs2005MNRAS}, although these are often affected by significant uncertainties due to spin noise modelling \citep{Parthasarathy2019}.
Direct distance constraints via timing parallax are typically limited to well-timed recycled pulsars. For most systems, distances are therefore estimated from DM combined with a Galactic electron density model, though this method has limited precision. The resulting distance uncertainties translate into substantial systematic errors in transverse velocity estimates for individual pulsars.
Recently, \citet{Ronchi2021ApJ} proposed a machine learning approach to infer the characteristic dispersion of a Maxwellian natal kick distribution using only proper motion measurements. Measuring proper motions for $\sim$2000 NSs with the SKAO telescopes would allow this method to constrain the underlying velocity dispersion with an uncertainty of approximately $60\;{\rm km\,s^{-1}}$.

High angular resolution interferometric imaging, in contrast to pulsar timing, can provide model-independent distance measurements and thus precise transverse velocity estimates for both recycled and non-recycled pulsars. Approximately 100 such measurements are currently available, with the majority obtained in recent decades through programs using the Very Long Baseline Array (VLBA), such as PSR$\pi$ \citep{Deller2019ApJ}.
This work showed that distances inferred from electron density models --- such as NE2001 \citep{Cordes2002astroph} and YMW16 \citep{Yao2017ApJ} --- can differ by factors of 3--5 compared to those derived from VLBI parallax measurements.

\begin{figure}[t!]
    \centering
    \includegraphics[width=0.47\textwidth]{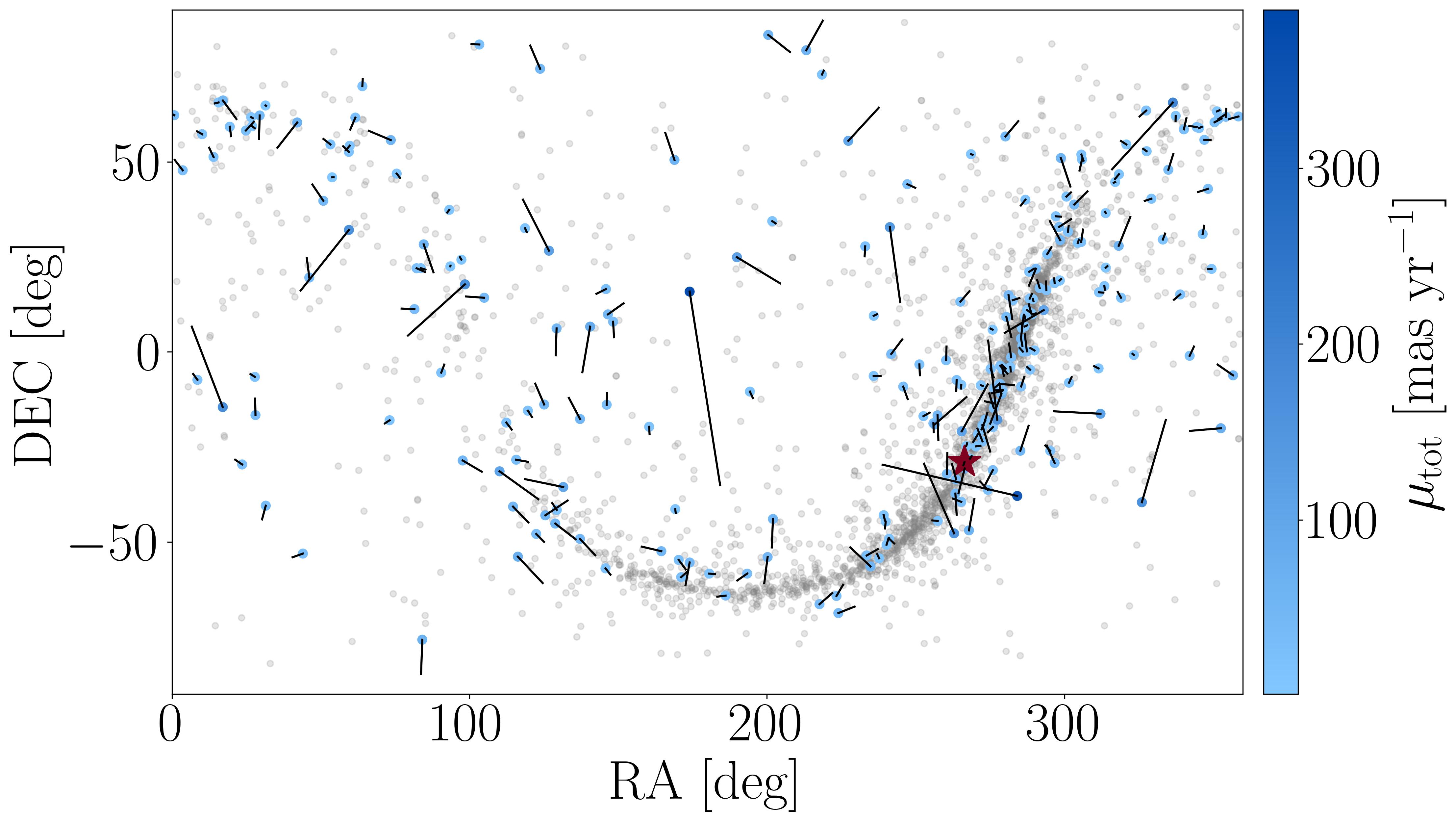}    
    \includegraphics[width=0.47\textwidth]{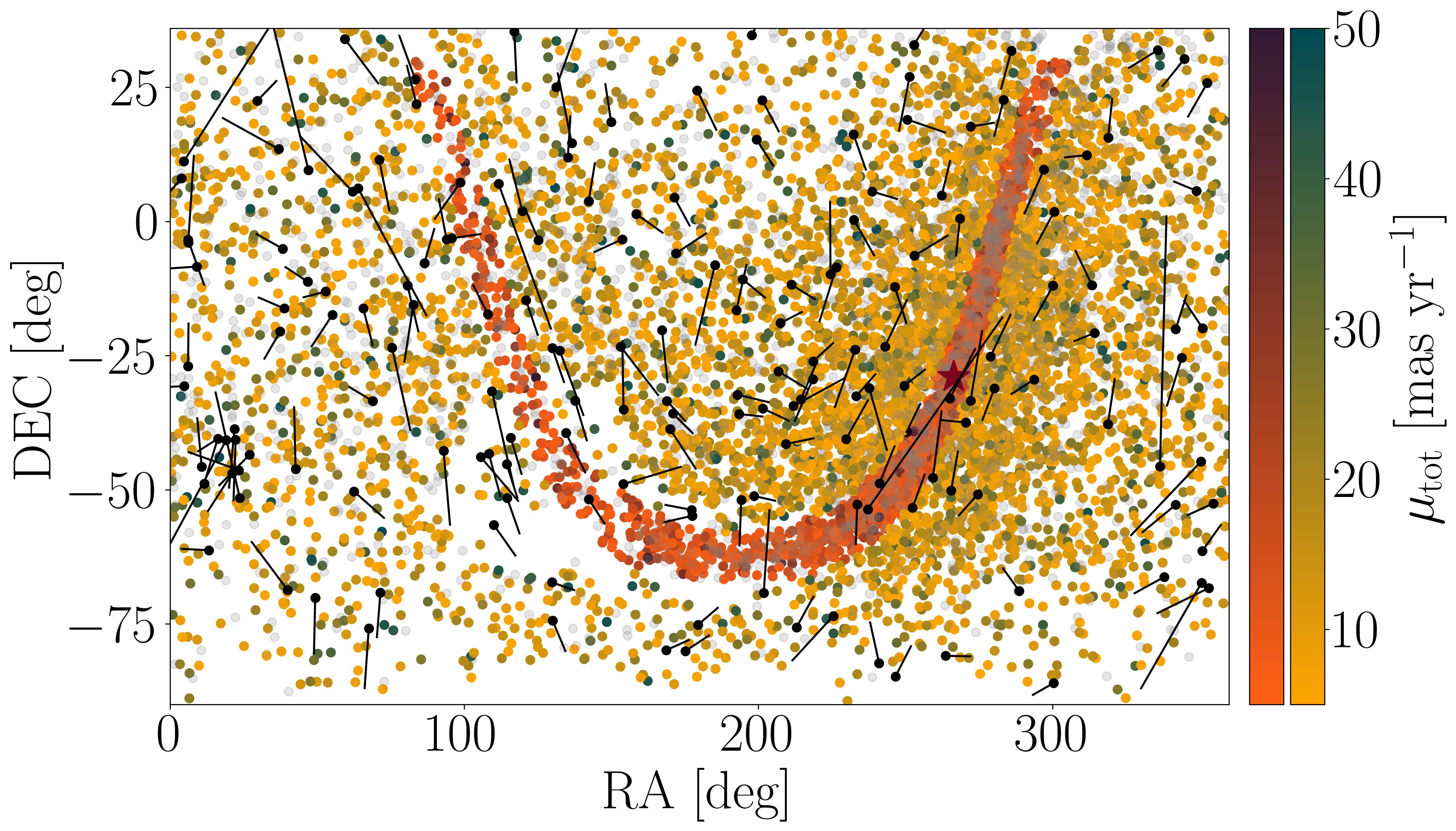}    
    \caption{Proper motions of radio pulsars. Top panel: All measured proper motions for standard radio pulsars listed in the ATNF Pulsar Catalogue v2.5.1; \citet[][]{Manchester2005}, \url{https://www.atnf.csiro.au/research/pulsar/psrcat/}. Bottom panel: Proper motions expected to be measured via timing with SKAO AA$^*$. The simulation corresponds to one realisation of Survey Option 3 with the evolutionary framework outlined in~\citet{Keane2025_SKA_Census}; the corresponding $P$-$\dot{P}$ diagram is shown in the left panel of Figure~\ref{fig:PPdot_SKA}. Grey dots indicate radio pulsars with $\mu_\mathrm{tot} < 5\;{\rm mas\,yr}^{-1}$, which will likely remain undetected by the SKA. Sources with larger proper motions are shown as coloured points according to the respective colour bars. We also show proper motion tracks extrapolated for the past $0.5\,$Myr. However, only the fastest pulsars with total proper motion $\mu_\mathrm{tot} > 50\;{\rm mas\,yr}^{-1}$ are highlighted with black lines in the bottom panel to reduce visual crowding.}
\label{fig:proper_motions_SKA}
\end{figure}

VLBI measurements of parallax and proper motion have been used to constrain the distribution of natal kicks \citep{Verbunt2017AA,Igoshev2020MNRAS}. The maximum likelihood techniques employed in these studies favour a bimodal velocity distribution, which can account for both fast and slow kicks. However, it remains unclear whether these two modes correspond to distinct physical mechanisms underlying natal kicks. Additional parallax and proper motion measurements are needed to refine these models and better connect them to the physics of SN explosions.

The SKAO telescopes will enable proper motion measurements for a large number of pulsars in the southern sky using pulsar timing techniques (see example in Figure~\ref{fig:proper_motions_SKA}). In addition, the SKAO will be integrated into VLBI networks to provide precise parallax and proper motion measurements for tens of radio pulsars in the southern hemisphere. These data will offer new constraints on the interstellar medium and contribute to refining the electron density model of the Milky Way.

\section{Long-period pulsars}
\label{sec:ULPs}

Until the late 2010s, the spin distribution of the observed isolated pulsar population ranged from tens of milliseconds for young systems up to around $10\,$s for the most magnetic NSs. 
As outlined in Section~\ref{sec:PPdot}, pulsar rotation periods in this range are relatively well understood as they are governed by the stars' initial spin periods and natal magnetic fields, and subsequent spin-down emission and magnetic field evolution \citep{Bates2014, Gullon2014, Gullon2015, Cieslar2020, Graber2024}. The upper limit in the period distribution has often been interpreted as the result of a highly resistive NS crust accelerating the field decay for stronger fields \citep{Pons2013}. However, given recent advances in search methods in the radio data from single dish and interferometric surveys, several slower, periodic, coherent and highly polarized signals have been discovered. In particular, three radio pulsars, PSR J1903$+$0433 \citep{Han2021}, PSR J0250$+$5854 \citep{Tan2018}, and PSR J0901$-$4046 \citep{Caleb2022} have been found to rotate at periods of 14\,s, 23\,s and 76\,s, respectively. While the former two sources could be accommodated within current model uncertainties, the discovery of a 76-second radio-loud pulsar began to challenge our understanding of isolated pulsar physics. On the one hand, it is unclear how isolated sources can attain periods $\gtrsim 50\,$s questioning current models of magneto-rotational evolution without invoking ad-hoc configurations of core-dominated magnetic fields. On the other hand, based on our existing knowledge of magnetospheric pair production and radio pulsar death lines, PSR J0901-4046's radio-loud nature is difficult to reconcile with standard emission models \citep{Zhang2000, Chen1993, Suvorov2023}.

\begin{figure}[t!]
\centering
\includegraphics[width=1.0\linewidth]{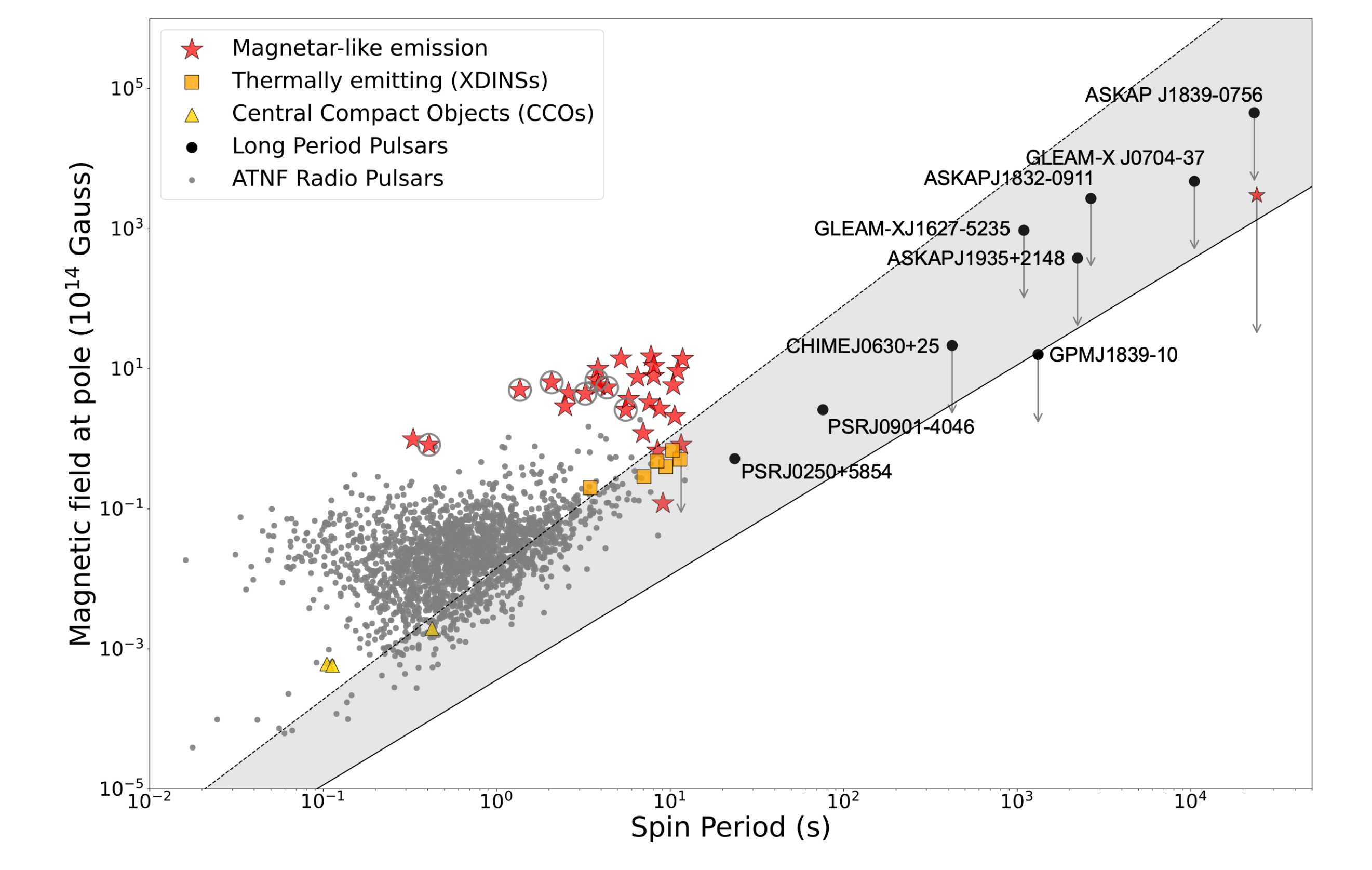}
\caption{Period-magnetic field diagram displaying the different classes of isolated NSs in addition to the periods and magnetic field limits of known long-period transients. The grey circles label the radio magnetars, and the rightmost red star is the CCO in RCW103, which has exhibited a magnetar outburst \citep{Rea2016}. The grey area represents a region of possible pulsar emission death lines based on various models  \citep{Wang2024}, with the upper limit representing a pure dipole model and the lower one showing a multipole model which is a factor of 10 stronger than the dipole.}
\label{fig:ppdot_LPTs}
\end{figure}

In recent years, both these issues have been further stretched by the discovery of several new radio sources with even longer spin periods ($\gg$ a few tens of seconds). As illustrated in Figure~\ref{fig:ppdot_LPTs}, to date, nine long-period transients have been discovered, with periods ranging between a few minutes to several hours. Most of these objects show transient radio activity, during which bursts with peak luminosities of $\sim$1\,mJ--50\,Jy appear periodically, with duty cycles ranging from $\sim$5-30\%. In one case, the periodic radio emission has been observed continuously for about 30 years (GPM\,J1839$-$10; \citealt{HurleyWalker2023}). In two systems, optical observations have identified the presence of a binary white dwarf system with an M-class star with a synchronized spin-orbit (ILT\,J1101$+$5521 and GLEAM-X\,J0704$-$36; \citealt{deRuiter2024, Hurley-Walker2024, Rodriguez2025}). For another source, transient X-ray emission has been detected during enhanced radio periodic activity \citep{Wang2024}. The radio emission in these sources shares many similarities with what is seen from radio pulsars (in particular, the high degree of linear polarization and the microstructures) and radio-emitting magnetars (see Section~\ref{sec:magnetars}) possibly suggesting that at least some of the long-period sources have a NS origin. 

Given the full range of observational properties, the nature of long-period radio transients might well be diverse, and the new class might not be composed entirely of NSs. A significant fraction might be a peculiar evolutionary stage of binary white dwarfs potentially related to radio-emitting white-dwarf systems such as AR Sco \citep{Marsh2016} and J1912$-$4410 \citep{Pelisoli2023}, or accreting magnetic cataclysmic variables \citep{Schreiber2021}. As we do not observe any radio pulsations from isolated white dwarfs \citep{Pelisoli2024} (despite many magnetic white dwarfs having periods in the observed range; \citealt{Rea2024}), the presence of a companion to produce the observed radio emission might be key for this scenario \citep{Buckley2017}. However, given the discovery of PSR J0901$-$4046 \citep{Caleb2022}, which is difficult to explain in the white-dwarf context, several of the newly discovered long-period sources may indeed be radio-emitting NSs. While angular momentum transfer from a fall-back accretion disk \citep{Chatterjee2000, Ho2017, Janka2022} could, in principle, provide a viable solution for driving significant spin-down in those NSs with the strongest magnetic fields \citep{Ronchi2022}, recent population analyses \citep[e.g.][]{Rea2024} suggest that the discovery rate of long-period sources in the NS-scenario alone cannot be easily reconciled with the fiducial Galactic core-collapse supernova rate of $\sim 2$ \citep{Rozwadowska2021}. In addition, the issue of radio pulsar emission at such long periods remains an open problem.

Despite several unanswered questions, the recent discovery of numerous long-period radio transients clearly indicates that previous time-domain radio surveys have been biased against a population of long-period NSs because they were not expected to be radio-loud. High-precision radio monitoring with the SKAO telescopes, combined with multi-wavelength observations, will be key to further exploring the distribution of long-period sources and their underlying nature. First, these studies will provide the most sensitive probes of the evolution of isolated NSs in our Galaxy, allowing us to examine the radio emission from objects close to the pulsar death valley and test the limits of theories of coherent radio emission \citep{Zhang2000, Chen1993, Suvorov2023}. In addition, the SKAO telescopes' unprecedented sensitivity may enable us to detect variability in timing residuals due to surrounding matter, enabling tests of the fallback disk evolutionary scenario. Finally, detection of X-ray or optical counterparts to these long-period transients will enable us to constrain models of the NS's magneto-rotational evolution. These aspects will ultimately have significant implications on the yield of pulsars with the SKAO.

\section{Observing modes and forecasts for SKAO telescope array assemblies AA* and AA4}
\label{sec:obsmodes}

Observations of NSs with the SKAO telescopes have great potential to significantly advance our understanding of all the different types of sources that have been discussed here, all across the NS family. 
The SKAO telescopes will be rolled out in so called Array Assemblies (AA), with the currently funded telescopes called AA* and the full design baseline called AA4.  
For SKA-Mid, AA* will consist of a total of 144 dishes (80 x 15 m diameter and 64 x 13.5 m diameter) with a maximum base line of 36 km, and AA4 will consist of 197 dishes (133 x 15 m diameter and 64 x 13.5 m diameter) with a maximum baseline of 150 km. 
For SKA-Low, AA* will consist of a total of 307 stations x 256 antennas, with a maximum base line of 74 km, and AA4 will consist of 512 stations x 256 antennas, also with a maximum baseline of 74 km. 
More details on array assemblies can be found on the SKAO website \footnote{e.g. https://www.skao.int/en/science-users/646/year-life-ska and https://www.skao.int/en/science-users/118/ska-telescope-specifications}. 

Both SKA-Mid and SKA-Low will be able to perform both pulsar search and pulsar timing, with a range of observing options and data products available. To maximise the scientific output, it will be essential to both find new sources in our Galaxy (and beyond) using the pulsar and single pulse search capabilities, and to set up an efficient pulsar timing programme to follow-up on new and previously known NSs using the pulsar timing capabilities. 
We outline below suggestions for how a pulsar survey could be set up, with forecasts for both isolated ordinary pulsars and for millisecond pulsars. We then discuss options for how to optimally set up a pulsar timing programme of known sources.

\subsection{Pulsar searches with AA* and AA4}

The pulsar search backend will be capable of searching for pulsars and fast transients in (quasi-)real time at both SKAO telescope sites. To get an estimate of the numbers of pulsars that may be detected with each telescope in the two Array Assemblies, we have run simulations both in snapshot mode using  PsrPopPy\footnote{https://github.com/samb8s/PsrPopPy} \citep{Bates2014} and in evolutionary mode following the methods described in \cite{Graber2024} and \cite{PardoAraujo2025}. The details of these simulations can be found in \cite{Keane2025_SKA_Census} as part of this special issue. A pulsar survey can be designed and optimised in a variety of ways, with different yields as outcome. One example from \cite{Keane2025_SKA_Census} shows that in a composite all-sky survey with SKA-Mid Band 2 focussing on the Galactic plane up to a Galactic latitude of 5$^\circ$ and SKA-Low covering the rest of the visible sky, the detected pulsar population would yield approximately 10,000 ordinary pulsars and 800 MSPs for AA*. The corresponding AA4 detection values are around 12,000 ordinary pulsars and 1,000 MSPs.

Thanks to the many simultaneous beams available in pulsar search mode, all these sources will be well localised instantly, which will in turn enables rapid follow-up with the SKAO telescopes and with other facilities both in the radio domain and in other wavelengths.

\subsubsection{Millisecond and binary pulsar surveys}

Currently, there are about 550 MSPs known in the Galactic field\footnote{https://www.astro.umd.edu/$\sim$eferrara/GalacticMSPs.html} (and an additional 345 pulsars in 45 Globular Clusters \footnote{https://www3.mpifr-bonn.mpg.de/staff/pfreire/GCpsr.html} at the time of writing). This number has grown rapidly over the last 10 years: the total number is three times higher than what was recorded in the 2015 SKA Science book \citep{Tauris2015SKA}. 
Our simulations suggest that surveys with the SKAO telescopes should be able to increase the number of MSPs in the Galactic field to $\sim$800 pulsars as part of AA* and $\sim$1000 pulsars in AA4 \citep{Keane2025_SKA_Census}. 

Many of the known MSPs are in binary systems, which enables a wide range of scientific studies, but also makes them more difficult to detect in the first place. 
The detectability of these pulsars degrades due to the orbital acceleration, and computationally intensive acceleration searches, such as those planned with the SKAO telescopes, are needed to detect them. With higher telescope sensitivity, shorter parts of the orbit can be sampled and still result in a detection, and hence some of the currently undetected MSP binary systems will be detectable with the SKAO telescopes. 
Some of these sources will be bright enough to follow-up with other radio telescopes, however it is likely that most sources will require the SKAO telescopes sensitivity to monitor them efficiently. 

As discussed in section \ref{sec:binaries}, many MSPs have been discovered by targetting unidentified Fermi $\gamma$-ray sources. 
In the future, additional targets may be found through  collaborations with other instruments. 
\cite{Korol2024} modelled the detectability of NS--white dwarf binaries with Laser Interferometer Space Antenna (LISA). They concluded that about $10^2$\,NS--white dwarf binaries with short orbital periods ($<$\,3 hours) can be discovered by LISA. A significant fraction of these sources are formed with NS recycling, i.e., the NS is a millisecond radio pulsar. The SKAO telescopes could detect some of these MSPs as a follow-up from a LISA discovery.

\subsubsection{Additional thoughts on surveys for long-period pulsars}

The SKAO telecopes could remove the observational selection bias for radio pulsation periods of more than a few seconds with respect to detecting and regular monitoring. 

For finding sources at the lower end of the long-period pulsar range, the use of a Fast Folding Algorithm (FFA) running on the pulsar survey data would be highly beneficial. An FFA search is more sensitive than traditional FFT searches, in particular in the case of long-period pulsars \citep[e.g.][]{Morello2020}. Running an FFA on the entire periodicity search range would be computationally expensive, but may be feasible by limiting the parameter space to the longest periods and would be an excellent addition to the pulsar search backend in the future. 

One of the key challenges to finding pulsars with long periods is the presence of red noise, due to the varying baseline of the time series. Techniques like subtraction of the incoherent beam power from each tied array beam would be highly advantageous in reducing the adverse effects of red noise when searching for long-period sources in particular. 

Furthermore, searching for transient radio emission in radio images at higher time resolution will allow for more and longer period discoveries. Along with single pulse searches in the time-domain, a fast imaging search is a proven technique for finding sources with long periods. This will be made possible by use of the Fast Imaging Pipelines at each SKAO telescope, which are planned to search for transients with a time scale of as short as 1 second in the fast-imaging products.

\subsection{Pulsar timing modes and regular monitoring} 
To enable all the science cases mentioned, it is not only necessary to find a large number of NSs, but all these sources also need to be monitored with pulsar timing to reveal their properties. 
With the large numbers of NS discoveries that are anticipated from the SKAO telescopes, it is essential to design an efficient strategy for follow-up observations and continuous monitoring of these pulsars. 
Some of the new discoveries may still be possible to observe with other radio telescopes, and collaboration between the SKAO and other radio telescopes should be highly encouraged, but many of the new sources are expected to fall below the detection threshold at many other facilities. To enable the science as discussed in this chapter, it is also crucial to regularly monitor many of the currently known NSs in addition to the new discoveries. 

For this to be feasible, it will be vital to make use of the multibeaming capability of the Pulsar Timing backend in combination with efficient use of telescope subarrays. A similar approach has been taken at the MeerKAT telescope over the last few years, through the  Thousand Pulsar Array  \citep[TPA,][]{johnston2020}\footnote{http://www.meertime.org/1000-pulsar-array.html} as part of the MeerTime project. 
The TPA was set up to regularly monitor at least 500 ordinary pulsars for a period of 5 years, and to obtain polarised pulse profiles for an additional 500 pulsars. The observation strategy is described in \cite{Song2021}, and discusses how the overall observing time can be reduced by the optimal use of MeerKAT's subarray capabilities. 
The TPA project's observation strategy first calculates the required integration length to reach a high-fidelity profile for each pulsar. It then compares the ratios of system noise to pulse-to-pulse variability to determine if subarraying would be beneficial if used for that particular pulsar. 
\cite{Song2021} also presents an example of observing 1000 pulsars with SKA-Mid, and shows that significant reductions in observing time can be achieved with careful consideration of subarrays. 
They find the highest time savings by allowing the 1000 pulsars to be observed in subarrays of different sizes, but fixed within an observing session. 
A similar analysis, using the latest array and telescope configurations for the various array assemblies, should be carried out closer to the start of SKAO operations to ensure optimal use of telescope time for pulsar timing observations.

Another aspect of ensuring efficiency of telescope time is the use of 'filler time'. Especially at the SKA-Low telescope, where other science goals at times may be restricted by e.g. the sun or the moon, short pulsar timing observations could be added in gaps in the schedule. It is also important to consider commensal observations where possible, as many sources may not need the full array or a large amount of computing resources. 
Making use of all these capabilities will help ensure the success of pulsar timing projects with the SKAO telescopes.

All these data, gathered with SKA-Low and SKA-Mid, will help us to study the different characteristics of radio-emitting neutron stars. These can be contrasted and compared to find differences and similarities. This will not only provide important insights into the emission processes of various types \citep[e.g.][]{Kramer2024}, but also shed light on potential evolutionary pathways connecting the different manifestations of NSs. This will help us to understand not only the specific sub-population, but also the aforementioned birth rate problem. A complete NS census with the SKA promises to provide a holistic view of NSs in the Milky Way and beyond.

\section*{Acknowledgments}
The authors would like to thank Aru Beri, Paolo Esposito, Fabian Jankowski, and Danny Vohl for providing helpful comments on the manuscript. V.~G. is supported by a UKRI Future Leaders Fellowship (grant number MR/Y018257/1). N.R. is supported by the European Research Council (ERC CoG No. 817661 and ERC PoC No. 101189496), and grants SGR2021-01269, ID2023-153099NA-I00, and CEX2020-001058-M.

\bibliographystyle{aasjournal}

\bibliography{chapter}

\end{document}